\documentclass[12pt]{emulateapj}
\pdfoutput=1   
\usepackage{natbib} 
\usepackage{epsfig} 
\usepackage{graphicx} 
\usepackage{color}
\usepackage{aas_macros} 
\usepackage{amssymb}
\usepackage{amsmath}
\usepackage[title]{appendix}

\newcommand{\enzo}{\texttt{Enzo~}}
\newcommand{\enzolat}{\texttt{Enzo-2.3}}

\newcommand{\kms} {km $\rm{s^{-1}}$}
 
\newcommand{\msolar} {$\rm{M_{\odot}}~$}
\newcommand{\msolarc} {$\rm{M_{\odot}}$}
\newcommand{\molH} {$\rm{H_2}$~}
\newcommand{\molHc} {$\rm{H_2}$}
\newcommand{\J} {$\rm{10^{-21}\ erg\ cm^{-2}\ s^{-1}\ Hz^{-1}\ sr^{-1}}$}

\submitted{Draft version \today}

\begin{document}
\title{The direct collapse of a massive black hole seed under the influence of 
an anisotropic Lyman-Werner source}

\author{John A. Regan$^{1}$\thanks{E-mail:john.regan@helsinki.fi}, Peter H. Johansson$^{1}$ \& John H. Wise$^{2}$}

\affil{$^1$ Department of Physics, University of Helsinki, Gustaf H\"allstr\"omin katu 2a,
FI-00014 Helsinki, Finland \\
$^2$ Center for Relativistic Astrophysics, Georgia Institute of Technology, 837 State Street, 
Atlanta, GA 30332, USA
}



\begin{abstract} 
\noindent The direct collapse model of supermassive black hole seed formation requires that the 
gas cools predominantly via atomic hydrogen. To this end we simulate the effect of an 
\textit{anisotropic} radiation source on the collapse of a halo at high redshift. The radiation 
source is placed at a distance of 3 kpc (physical) from the collapsing object and is set 
to emit monochromatically in the center of the Lyman-Werner (LW) band. The LW radiation 
emitted from the high redshift source is followed self-consistently using ray tracing 
techniques. Due to self-shielding, a small amount of \molH is able to form at the very 
center of the collapsing halo even under very strong LW radiation. Furthermore, we find that 
a radiation source, emitting $> 10^{54}\ (\sim10^3\ \rm{J_{21}})$ photons per second is 
required to cause the collapse of a clump of $\rm{M \sim 10^5}$ \msolarc. The resulting 
accretion rate onto the collapsing object is $\sim 0.25$ \msolar $\rm{yr^{-1}}$. 
Our results display significant differences, compared to the isotropic radiation field case, 
in terms of \molH fraction at an equivalent radius. These differences will significantly effect 
the dynamics of the collapse. With the inclusion of a strong anisotropic radiation source, the 
final mass of the collapsing object is found to be $\rm{M \sim 10^5}$ \msolarc. This is consistent 
with predictions for the formation of a supermassive star or quasi-star leading to a 
supermassive black hole. 
\end{abstract}



\section{Introduction} 
\noindent Observations of supermassive black holes (SMBH) at redshifts greater than 
$z\gtrsim 6$ \citep{Fan_2004, Fan_2006, Mortlock_2011,Venemans_2013} has led to difficulties
in understanding how such large objects could have formed so early in the Universe. 
The most obvious route is via a Population III (Pop III) star which after its initial 
stellar evolution, collapses and forms a stellar mass black hole which can then grow 
to become a SMBH by a redshift of $z\sim 6$. However, a number of authors 
\citep[e.g.][]{Alvarez_2009, Milosavljevic_2009, Johnson_2013b, Jeon_2014}
have shown that this scenario suffers from several severe limitations.
Most pertinent is the fact that in order for a stellar seed black hole to 
grow to a supermassive size by a redshift of $\rm{ z \sim 7}$ it is necessary for the 
seed to grow at the Eddington limit for almost the entire time. \\
\indent A compelling solution is to start with a significantly larger seed mass than the 
mass now advocated for the first stars. Recent simulations of Pop III collapse has put their 
mean mass at below $M_* \lesssim 100$ \msolar 
\citep{Greif_2011, Greif_2012, Stacy_2012, Turk_2012, Hirano_2014}. If instead 
we start with a much larger seed mass the restrictions on the growth rate are eased significantly. 
The so-called direct collapse model leads to initial masses of between $10^4$ and $10^6$ \msolarc.
Initial work on the method began with pioneering work by \cite{Loeb_1994} and \cite{Eisenstein_1995b}
who considered the direct collapse of gas into a massive black hole seed that could then 
power the quasars observed at high redshift. Further work in recent years 
\citep{Begelman_2006, Lodato_2006, Wise_2008, Volonteri_2005b, Volonteri_2008, Volonteri_2010, 
Johnson_2011,Regan_2009b, Regan_2009, Agarwal_2013,  Agarwal_2014b, Agarwal_2014, Latif_2013c, 
Regan_2014a, Johnson_2013b, Johnson_2014} has led to a growing appreciation 
that the direct collapse method is a viable alternative.  \\
\indent In order to create the conditions in 
which a massive seed may form, a halo that can support atomic hydrogen cooling, having a virial 
temperature $T_{\rm vir} \ga 10^4$ K, is required which is free of metals, dust
and $\rm{H_2}$. Metals, dust and \molH would enhance cooling to the point where the gas would 
fragment into small clumps and eventually form a star of mass less than $\rm{M_* \sim 100}$ 
\msolarc. At early times in the Universe we expect the contribution from both metals and dust to be 
negligible, \molH can however readily form in regions of moderate to high gas density. \\
\indent A large seed mass can only form if the corresponding Jeans mass of the collapsing object
remains high. This can be achieved if the gas temperature stays close to the virial temperature 
of the halo, cooling by neutral hydrogen allows the gas to cool to approximately 
$T\sim 6000 \ \rm K$ and facilitates the collapse to a large seed mass. 
Early numerical work by \cite{Wise_2008} and \cite{Regan_2009} showed that in the absence 
of \molH the gas could cool isothermally and collapse to form a disk-like structure with a mass 
of a few times $10^4$ \msolarc, which could then go on to form a super-massive star 
\citep[e.g.][]{Inayoshi_2014, Inayoshi_2014b}, a quasi-star \citep{Begelman_2006, Ball_2011}, 
or a  dense stellar cluster \citep[e.g.][]{Gurkan_2004, Gurkan_2006}. \\
\indent In assuming the absence of \molH previous studies have generally assumed that the 
\molH can be efficiently dissociated by a nearby source peaking in the Lyman-Werner (LW) band 
(11.2-13.6 eV). LW photons dissociate \molH by exciting electrons to higher energy levels 
resulting in the breakup of the molecule. A number of authors have examined such a scenario, 
both using semi-analytic models \citep{Dijkstra_2008, Dijkstra_2014} and using 
numerical simulations \citep{Shang_2010, Latif_2014a, Latif_2014b, Agarwal_2013, Agarwal_2014, 
Johnson_2014}. However, in all of the above numerical work the authors have assumed an 
isotropic background.  At high redshift when local sources dominate over the LW background, this is 
likely to be an incorrect assumption given the highly anisotropic nature of early structure 
formation and the evidence accumulated for a extended period of reionisation 
\cite[e.g.][]{Fan_2006}.\\ \indent In this paper we instead model a highly anisotropic source, 
ignoring the effects of a possible isotropic LW background. We place a source at a distance of 
3 kpc from a collapsing mini-halo and turn the source on before the mini-halo collapses due 
to \molH cooling. We ignore the effect of a LW background and instead concentrate on the effect 
of the nearby source only, the high redshift of the collapse ($z > 20$) means that any LW 
background at this redshift is likely to be patchy.  Previously, \citet{Shang_2010} and 
\citet{Agarwal_2014} considered local \molH self-shielding effects, which is intrinsically an 
integrated property and should depend on the non-local environment.  Improving upon this local 
approximation, we use ray-tracing to calculate the dissociating effects of a LW source 
self-consistently. We run several realisations, using the same halo in each case but varying the 
flux intensity. The goal of this work is to analyse the effect of an \textit{anisotropic} source 
on the formation of a massive black hole seed and to determine the intensity of the 
\textit{anisotropic} flux required to ensure the halo remains \molH free. The model simulates 
the effect of a close halo pair, believed to be required to provide the necessary LW flux 
\citep{Dijkstra_2008, Visbal_2014b}. \\
\indent The paper is laid out as follows: in \S \ref{Sec:NumericalSetup} we describe the 
numerical approach used, in \S \ref{Sec:J21} we relate the flux from an anisotropic flux to that 
from an isotropic field,  in \S \ref{Sec:Results} we describe the results of our 
numerical simulations, in \S \ref{Sec:Isotropic} we compare our anisotropic results against
simulations using an isotropic radiation field, in \S \ref{Sec:Discussion} we analyse the 
results and in \S \ref{Sec:Conclusions} we present our conclusions.  
Throughout this paper we  assume a standard $\Lambda$CDM cosmology with the following parameters 
\cite[based on the latest Planck data]{Planck_2013a}, $\Omega_{\Lambda,0}$  = 0.6817, 
$\Omega_{\rm m,0}$ = 0.3183, $\Omega_{\rm b,0}$ = 0.0463, $\sigma_8$ = 0.8347 and $h$ = 0.6704. 
We further assume a spectral index of primordial density fluctuations of $n=0.9616$.\\


\begin{table*}
\centering

\scriptsize

\begin{tabular}{ | l | l | l | l | l | l | l | l |l |l| l |}

\hline

\em{\rm{Sim}}$^a$ & \textbf{\em $\rm{Halo\ Description}^{b}$}
& \textbf{\em $\rm{Source\ Flux}^{c}$} & \textbf{\em $z_{\rm end}^{d}$} & \textbf{\em $M_{200}^{e}$} 
& \textbf{\em $R_{200}^{f}$} & \textbf{\em $V_{200}^{g}$}  & \textbf{\em $T_{\rm vir}^{h}$} 
& \textbf{\em $n_{\rm H, \rm max}^{i}$} & \textbf{\em $T_{\rm core}^{j}$} & \textbf{\em $M_{\rm core}^{k}$} \\ 
\hline 
1050 & Mini-Halo & $1 \times 10^{50}$ & 29.15 & 1.08 $\times 10^6$ & 0.11 & 6.64 & $1585$ 
& $6.08 \times 10^8$ & 445 & $1.42 \times 10^3$ \\
1051 & Mini-Halo & $1 \times 10^{51}$  & 23.82 & 3.20 $\times 10^6$ & 0.18 & 8.64 & $2684$ 
& $4.14 \times 10^8$ & 835 & $4.46 \times 10^2$  \\
1052 & Atomic Cooling Halo  & $1 \times 10^{52}$ & 22.01 & 2.86 $\times 10^7$ & 0.41 
& 17.26 & $10725$ & $3.93 \times 10^7$ & 490 & $4.30 \times 10^3$   \\
1054 & Atomic Cooling Halo & $1 \times 10^{54}$  & 21.26 & 5.65 $\times 10^7$ & 0.54 
& 21.29 & $16329$ & $2.47 \times 10^{7}$ & 850 & $7.63 \times 10^3$ \\ 
1056 & Atomic Cooling Halo & $1 \times 10^{56}$  & 21.27 & 5.58 $\times 10^7$ & 0.53 
& 20.21 & $16203$ & $2.62 \times 10^{7}$ & 4298 & $5.5 \times 10^3$ \\ 
1058 & Atomic Cooling Halo & $1 \times 10^{58}$  & 21.22 & 5.78 $\times 10^7$ & 0.54 
& 21.44 & $16557$ & $1.64 \times 10^{7}$ & 5223 & $9.05 \times 10^3$ \\ 
\hline

\end{tabular}
\normalsize
\caption{Simulation and halo details}
\parbox[t]{0.9\textwidth}{\textit{Notes:} The above table contains the simulation name$^a$, 
  the halo description (either 
  a mini-halo cooled predominantly by \molH or an atomic cooling halo cooled predominantly 
  by H)$^b$,  the source flux in photons per second$^{c}$, the redshift$^{d}$ on reaching the 
  highest refinement level, the total mass$^{e}$ (gas \& dark matter) at the virial 
  radius\footnotemark[2] at $z_{\rm end}$ [$M_{\odot}$], the virial radius$^{f}$ [kpc], the virial 
  velocity$^g$ $(v_{\rm vir}=\sqrt{GM_{\rm vir}/r_{\rm vir}})$ [\kms], the virial temperature$^{h}$ [K], 
  the maximum gas number density$^{i}$ in the halo [$\rm{cm^{-3}}$], the temperature$^{j}$ at the 
  core\footnotemark[3] of the halo [K] and the enclosed gas mass$^{k}$ within the core of the 
  halo [$M_{\odot}$]. All units are physical units, unless explicitly stated otherwise.}

\label{Table:SimDetails}

\end{table*}



\begin{table*}
\centering

\begin{tabular}{ | l | l | l | c |}
\hline 
\textbf{\rm{Sim}$^a$} & \textbf{\rm Photons Per Second$^{b}$} & 
\textbf{\rm Flux at Max Density$^{c}$} & \textbf{\em{$\rm{J^{d}}$}} \\ 
\hline 
1050 & $10^{50}$ & 6.44 $\times 10^4$ & 1.36 $\times 10^{-1}$   \\
1051 & $10^{51}$ & 6.44 $\times 10^5$ & 1.36 $\times 10^0$    \\
1052 & $10^{52}$ & 6.44 $\times 10^6$ & 1.36 $\times 10^1$    \\
1054 & $10^{54}$ & 6.44 $\times 10^8$ & 1.36 $\times 10^3$    \\ 
1056 & $10^{56}$ & 6.44 $\times 10^{10}$ & 1.36 $\times 10^5$  \\ 
1058 & $10^{58}$ & 6.44 $\times 10^{12}$ & 1.36 $\times 10^7$  \\ 
\hline 

\end{tabular}
\caption{Relationship between radiation source and incident flux}
\parbox[t]{0.9\textwidth}{\textit{Notes:} The above table contains the simulation name$^a$, 
  the photons emitted per second$^{b}$ at the source, 
  the flux$^{c}$ at the point of maximum density in units of photons per second 
  per cm$^2$ and the spectral flux$^{d}$ at the point of maximum density in units of $J_{21}$.
  The values are calculated when the source initially turns on. }

\label{Table:FluxTable}
\end{table*}


\section{Numerical Setup} \label{Sec:NumericalSetup}
\noindent We have  used the publicly available adaptive mesh refinement
(AMR) code \texttt{Enzo}\footnote{http://enzo-project.org/}. The code has matured 
significantly over the last few years and as of July 2013 is available as 
version \enzolat \ with ongoing development of the code base among a wide range of developers. 
Throughout this study we use \enzo version 2.3\footnote{Changeset 0aa82394b23d+} with some 
modifications to the Radiative Transfer component (see \S \ref{RP}).\\
\indent \enzo was originally developed  by Greg  Bryan  and
Mike  Norman at  the University of  Illinois \cite[]{Bryan_1995b, 
Bryan_1997, Norman_1999, OShea_2004, Enzo_2014}. The gravity solver 
in \enzo uses an $N$-Body adaptive particle-mesh technique \citep{Efstathiou_1985,  
Hockney_1988, Couchman_1991} while the hydrodynamics are evolved using 
the piecewise parabolic method combined with a non-linear Riemann
solver for shock capturing. The AMR methodology allows for 
additional finer meshes to be laid down as the simulation runs to enhance the resolution
in a given, user defined, region. \\
\indent The  Eulerian AMR  scheme  was  first
pioneered by \cite{Berger_1984} and \cite{Berger_1989} to solve the  hydrodynamical 
equations for an ideal gas. \cite{Bryan_1995b} successfully ported the mechanics of the AMR
technique to cosmological simulations. In addition to the AMR there are also modules 
available which compute the radiative cooling of the gas together with a multispecies 
chemical reaction network. Numerous chemistry solvers are now available as part of the 
\enzo package. For our purposes we use the nine species model which includes:
${\rm H}, {\rm H}^+, {\rm He}, {\rm He}^+,  {\rm He}^{++}, {\rm e}^-$,
$\rm{H_2, H_2^+ and\ H^-}$. 
We allow the gas to cool radiatively during the course of the simulation  
Furthermore, we use the formation rates and collisional dissociation rates 
from \cite{Abel_1997} with the exception of the \molH collisional dissociation rate where we 
adopted the rates from \cite{Flower_2007}.\\
\indent For our simulations the maximum refinement level is set to 18. The maximum 
particle refinement level is set as the default \enzo value (i.e. equal to the 
maximum grid refinement level). We initially ran convergence 
tests to determine the most appropriate value for the maximum particle refinement level 
and found that as we lowered the maximum level the results became unconverged. We therefore
chose the default \enzo value. The simulations are 
allowed to evolve until they reach this maximum refinement level at which point they are terminated.
Our fiducial box size is 2 $\rm{h^{-1}}$ Mpc comoving giving a maximum comoving resolution of 
$\sim  6\ \times 10^{-2}\ \rm{h^{-1}}\ $ pc. 
Initial conditions were generated with the ``inits'' initial 
conditions generator supplied with the \enzo code. 
The nested grids are introduced at the initial conditions stage. 
We have first run exploratory dark matter (DM) only simulations with coarse resolution, 
setting the maximum refinement level to 4. These DM only simulations have a root
grid size of $256^3$ and no nested grids. 
For these simulations we originally ran 150 DM simulations and identified the most massive peak 
at a redshift of $\rm{z = 30}$. Using the initial conditions seed from the DM only simulations 
we then reran the simulation with the hydrodynamic component. We also included three levels of 
extra initial nested grids around the region of interest, as identified from the coarse
DM simulation. This led to a maximum effective resolution of $1024^3$.
The introduction of nested grids is accompanied by a corresponding increase in the DM resolution 
by increasing the number of particles in the region of interest. The DM particle resolution 
within the highest resolution region is $8.301 \times 10^2$ \msolarc. Within this highest 
resolution region we further restrict the refinement region to a comoving region of size 
$128 \, \rm{h^{-1}}$ kpc around the region of interest so as to minimise the computational 
overhead of our simulations. We do this for all of our simulations. The total number of 
particles in our simulation is 4,935,680, with $128^3$ of these  in our highest resolution
region. The grid dimensions at each level at the start of the simulations are as follows:
L0[$128^3$],
L1[$64^3$],
L2[$96^3$],
L3[$128^3$].\\
\indent Furthermore, the refinement criteria used in this work were based on three 
physical measurements: (1) The dark matter particle over-density, (2) The baryon over-density 
and (3) the Jeans length. The first two criteria introduce additional meshes when the over-density 
(${\Delta \rho \over \rho_{\rm{mean}}}$) of a grid cell with respect to the mean density exceeds 
3.0 for baryons and/or DM. Furthermore, we set the \emph{MinimumMassForRefinementExponent} 
parameter to $-0.1$ making the simulation super-Lagrangian and therefore reducing the threshold for
refinement as higher densities are reached \cite[]{OShea_2008}. For the final criteria we set 
the number of cells per Jeans length to be 16 in these runs. Recent studies \citep{Federrath_2011, 
Turk_2012,Latif_2013b} have shown that a resolution of greater than 32 cells per Jeans length 
may be required to fully resolve fragmentation at very high resolutions. However, at the 
resolution probed in this study this is unlikely to be a concern.
\footnotetext[2]{The virial mass is defined as 200 times the mean density of the 
Universe in this case.}
\footnotetext[3]{The core is here defined as the region within 1 parsec of the
point of maximum density.}
\begin{figure*}
  \centering 
  \begin{minipage}{175mm}      \begin{center}
    \centerline{\includegraphics[width=14cm]{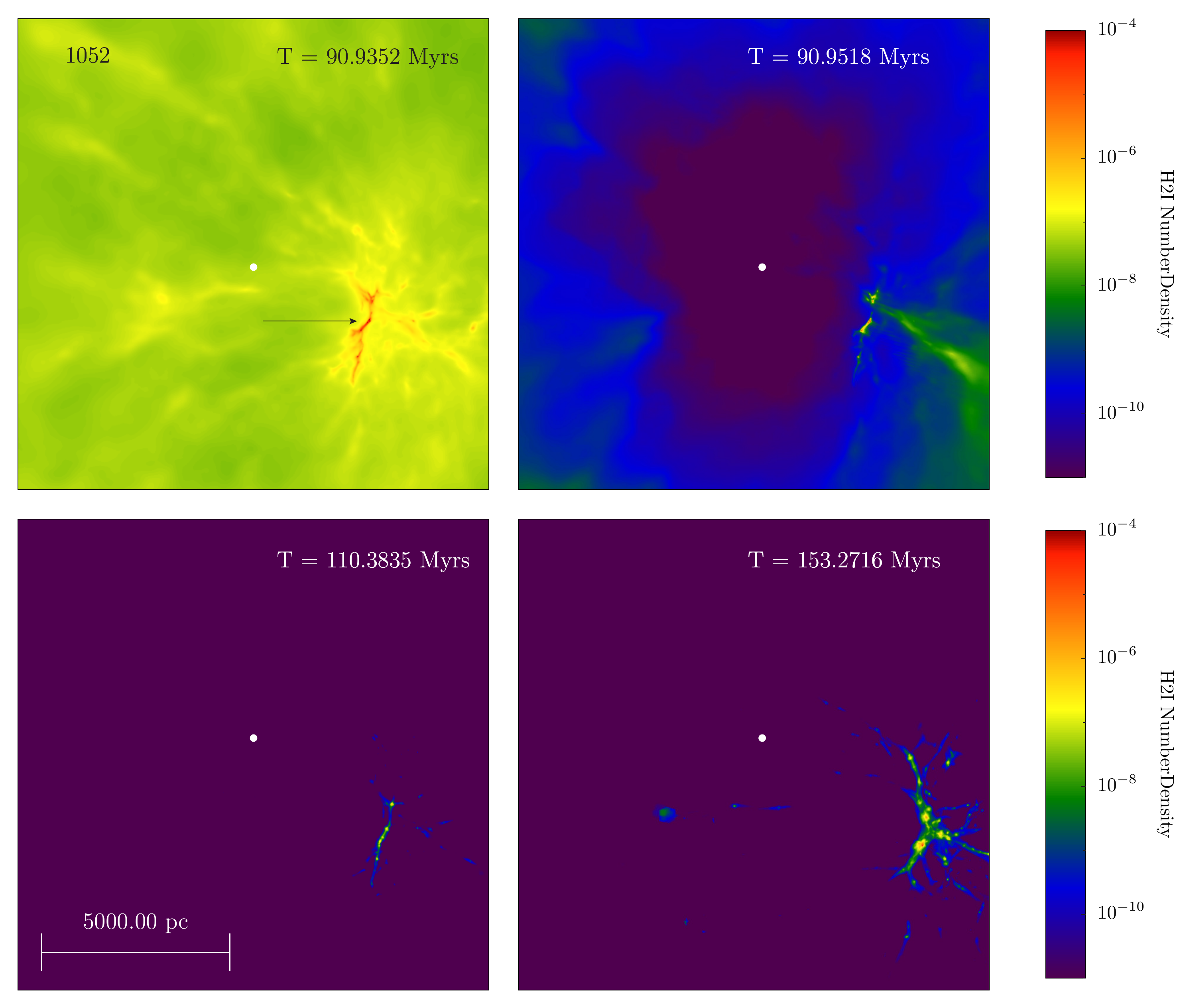}}
    \caption[]
    {\label{InitialSource}
      The neutral H2I density evolution shown, in projection, for simulation 1052 
      for illustration. The source is switched on at z = 32, which corresponds to T = 90.9352 
      Myrs - top left panel.The source is represented by the filled white circle.
      The arrow in the top left panel points at the center of 
      maximum density of the collapsing halo of interest. As the photons from the 
      source travel outwards they dissociate \molHc. The top right hand panel shows 
      the H2I number density approximately $10^5$ yrs later. At this stage the 
      density of \molH has been reduced by several orders of magnitude. As the time 
      evolution continues the flux from the source keeps the \molH levels strongly 
      suppressed. However, as the simulation proceeds the gravitational collapse and 
      subsequent increase in density of atomic hydrogen enables the formation rate of 
      \molH in the collapsing structure to out-pace that of the dissociating background.
      The \molH formation rate is driven by the reaction H$^-$ + H $\rightarrow$ \molH + e.
      In the bottom right panel the growth of the halo and the increased density of $\rm{H_2}$,
      in the face of a dissociating source, is clearly visible.
    }
   \end{center} \end{minipage}
\end{figure*}


\subsection{Halo Selection}
\noindent Only a single halo is used in this study that has a mass $M \simeq 10^6 M_\odot$ 
at $z = 30$ and grows to $M \simeq 6 \times 10^7 M_\odot$ by $z = 21$. The simulation is rerun multiple times 
with different radiation parameters, as detailed in Table \ref{Table:SimDetails} but 
the initial conditions are unchanged for each run. The halo was originally identified in a 
previous study \citep{Regan_2014a}. It corresponds to a very rare peak in the 
linearly extrapolated density field. In terms of the rms fluctuation amplitude, $\nu = 4.2$, 
where $\nu$ is defined as
\begin{equation}
\nu = {\delta_c \over \sigma(M) D(z)}
\end{equation}  
where $\delta_c \approx 1.686$ is the threshold over-density for a spherical collapse 
\citep{Gunn_1972}, $D(z)$ is the growth factor and $\sigma(M)$ is the mass fluctuation 
inside a halo of mass M. The mass fluctuation is given by 
\begin{equation}
\sigma^2(M) = \int {k^2 \over {2 \pi^2}} P(k) W(kR) dk
\end{equation}  
\noindent where the integral is over the wavenumber $k$, $P(k)$ is the power spectrum and 
$W(kR)$ is the top hat window function. In this context $\nu = 4.2$ corresponds to a very rare 
halo. A $\nu = 4.2$ peak corresponds to a host halo with a mass of $\rm{M} \approx 1 \times 10^{12}$ 
\msolar at $\rm{z \sim 6}$. It was convenient in this case to look for a halo collapsing early, 
and rapidly, to alleviate the computational demands set by the radiative transfer module. 
In addition, the very high redshift of the collapse $(z\sim 20-30)$ in this case strengthens 
our assumptions of a negligible global LW background as well as the absence of metals and dust. 
Furthermore, these rare density peaks are the most likely progenitors of the $z \ga 6$ quasar 
hosts \cite[e.g.][]{Costa_2014}.

\subsection{Radiative Particles} \label{RP}
\noindent The simulations conducted in this paper used a massless radiation source particle. 
We added this feature to the stable version of the \enzo code. In order to
complete the modification the new particle type was coupled together with the 
radiative transfer module so that the particle became a source particle capable of 
producing a LW flux of a given flux density. The active particle is not created 
on the fly as it does not result from the collapse of gas or any other physical mechanism. 
The code is stopped at a predefined point in time, the particle's coordinates are supplied 
and the particle is inserted into the code using a simple input file. The particle data is 
read by the code and is recognised as a radiation source particle. We choose the current 
approach as it gives us maximum flexibility in terms of 
where we put the source particle relative to the halo of interest. We now describe the 
radiative transfer setup used in this work.

\subsection{Radiative Transfer Setup} \label{Sec:RadTransferSetup}
\noindent The \molH dissociating radiation emitted by the massless source particle is propagated 
with adaptive ray tracing \citep{AbelWandelt_2002, WiseAbel_2011} that is based on the 
HEALPix framework \citep{Gorski_2005}. The radiation field is evolved at every hydrodynamical 
timestep of the finest AMR level. The \molH dissociation that occurs at each timestep couples to 
the hydrodynamical component self-consistently.  The photons travel at infinite speed through the 
simulation at each timestep with the photons halted when one of the following conditions is met:
\begin{enumerate}
\item
  The photon travels 0.7 times the simulation box length
\item
  The photon flux is almost fully absorbed ($>99.9\%$) in a single cell. 
\end{enumerate}
Photons are therefore traced, at each hydrodynamic timestep, through the entire region of interest.
The instant light propagation is motivated by the fact that the dynamical time is long compared
to the light propagation timescale.
\begin{figure*}[t!]
  \centering 
  \begin{minipage}{175mm}      \begin{center}
      \centerline{
        \includegraphics[width=9cm]{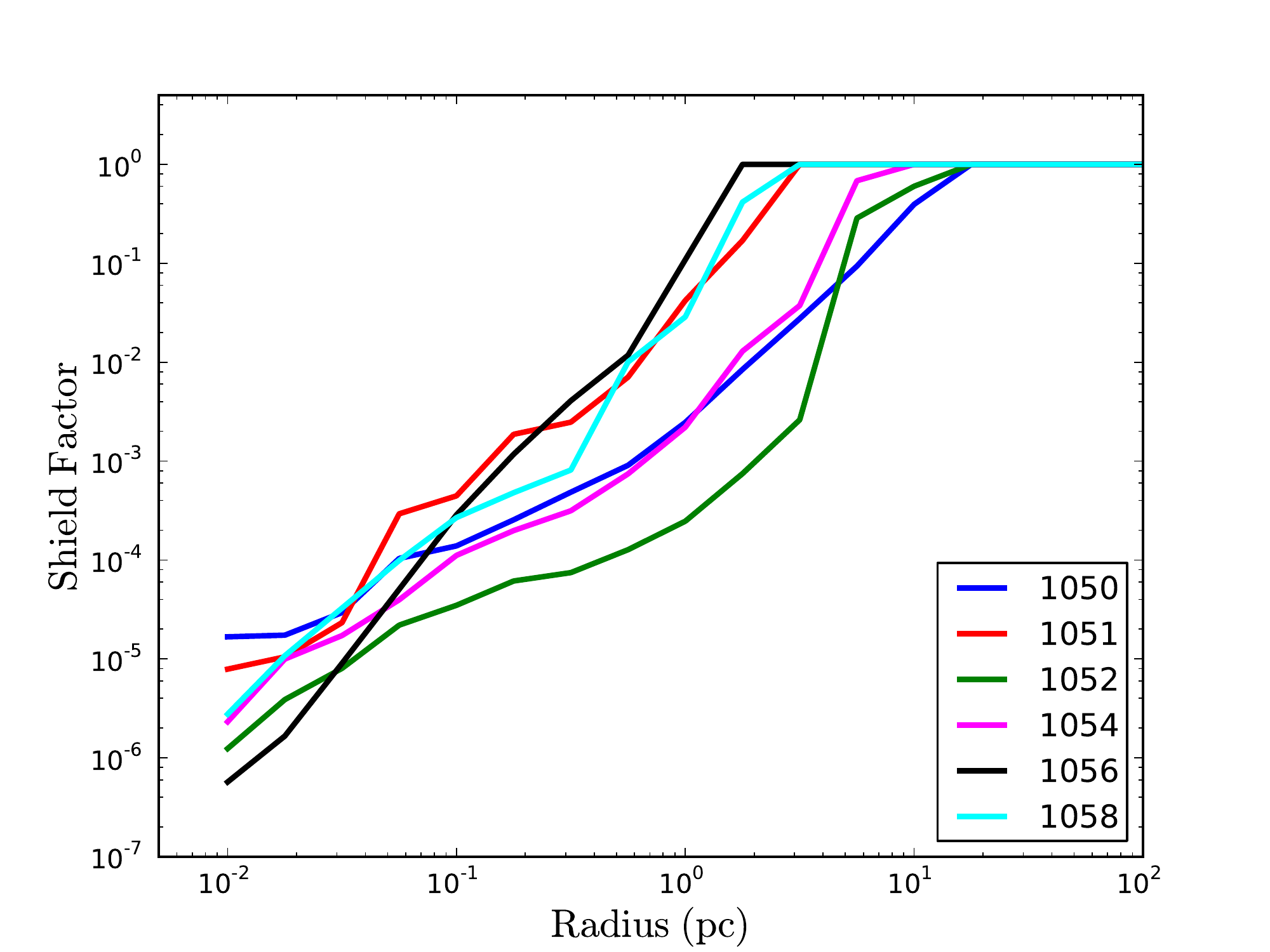}
        \includegraphics[width=9cm]{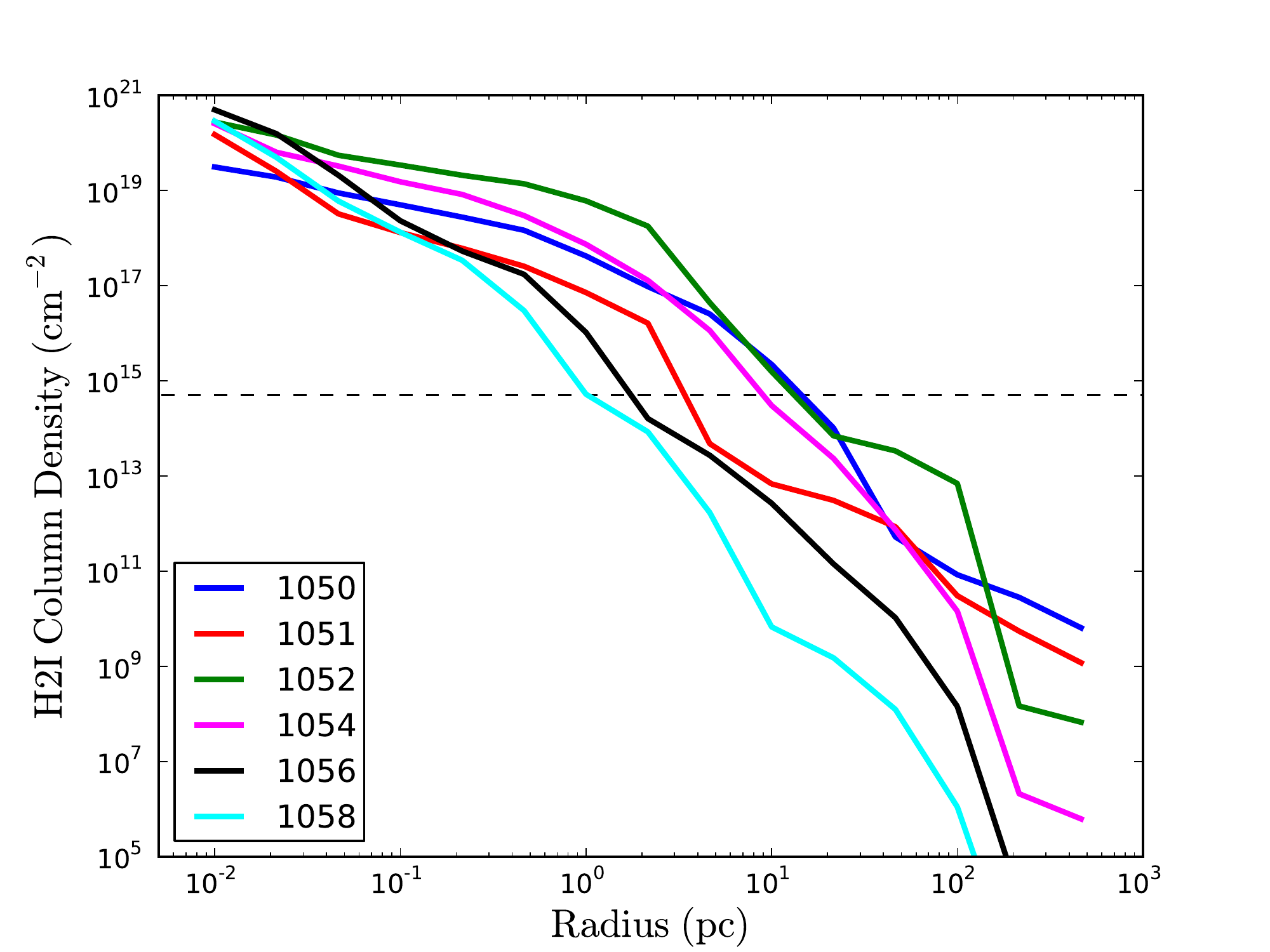}}
        \caption[]
        {\label{ShieldColumn}
          The left panel shows the shielding factor calculated at the moment each 
          simulation reaches the highest level of refinement (18). To 
          calculate the shielding factor 500 sightlines are drawn from the 
          source to an area surrounding the point of maximum density. The average
          over all sightlines is then used. The shielding
          factor is determined using the formula given in WG11 (and reproduced here
          as equation \ref{DB37}) which depends 
          both on the column density of the gas and its temperature. The right hand 
          panel shows the \molH column density found along the same sightlines as the 
          shielding factor. The dashed line indicates the column density above which 
          the \molH is assumed to be optically thick (see eqn \ref{DB37}). 
          Note that the HI column densities would be a factor of approximately 
          $\sim 10^3$ higher in the central regions than the \molH column densities shown here.
        }
      \end{center} \end{minipage}
  \end{figure*}

Photons dissociate \molH as they travel outwards (see Figure \ref{InitialSource}) from the source.
We use an average cross-section for \molH dissociation of $3.71 \times 10^{-18}$ cm$^2$ to calculate
the dissociation rate as photons pass through the gas. The medium through which the photons 
travel is assumed to be optically thin below a \molH column density of 
$5 \times \rm{10^{14}\ cm^{-2}}$. Above this limit the self-shielding approximation
taken from \cite{Wolcott-Green_2011}, hereafter WG11, is used. The shielding approximation 
is based on earlier work by \cite{Draine_1996}. In the high column density regime \molH 
is assumed to be optically thick and the 
dissociation rate is calculated using a fitting function. The fitting function is given by 
\begin{equation} \label{DB37}
\begin{split}
\rm{Shield\ (N_{H_2})}\ &= {0.965 \over (1 + X/b_5)^{\alpha}} \\ & + {0.035 \over (1 + X)^{0.5}}
\times \rm{exp}[-8.5 \times 10^{-4} (1 + X)^{0.5}]
\end{split}
\end{equation}
where $\alpha$ is set to be 1.1 in this study, $X \equiv N_{H_2} / (5 \times 10^{14}\ \rm{cm^{-2}})$, 
$\rm{b_5 \equiv b/(10^5\ cm\ s^{-1})}$ and b is the usual Doppler parameter in this case. 
This fitting function is used to accurately account for the self-shielding of \molH
from dissociating radiation which occurs at column densities above $5 \times 10^{14}\ \rm{cm^{-2}}$; 
see WG11 for more details.
The use of a self-shielding approximation is required in simulations using radiative transfer
techniques at these scales so as to make the simulation computationally viable with the expected
errors from using such fits expected to be small \citep{Draine_1996}. \\
\indent The massless source particles used in our simulations are all monochromatic, emitting 
radiation at the center of the Lyman-Werner band only, the energy of the photons is set to be 
$12.8$ eV  $(\lambda = 96.9 \ \rm{nm})$ in all cases. The details of each simulation is given in 
Table \ref{Table:SimDetails}. The name of each simulation gives the source flux,
for example simulation 1052 has a source flux of $10^{52}$ photons per second.

\begin{figure*}[t!]
  \centering 
  \begin{minipage}{175mm}      \begin{center}
      \centerline{
        \includegraphics[width=9cm]{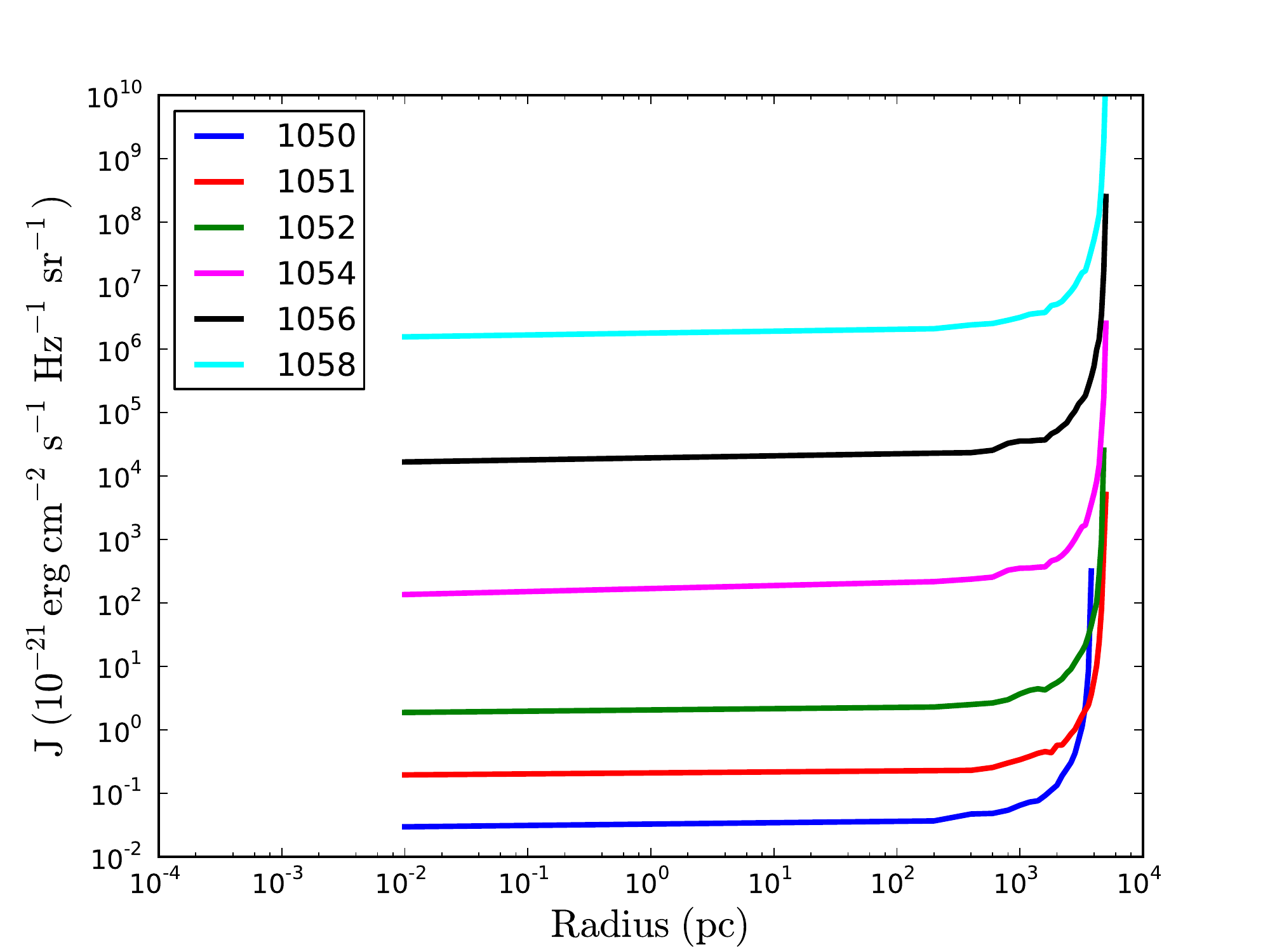}
        \includegraphics[width=9cm]{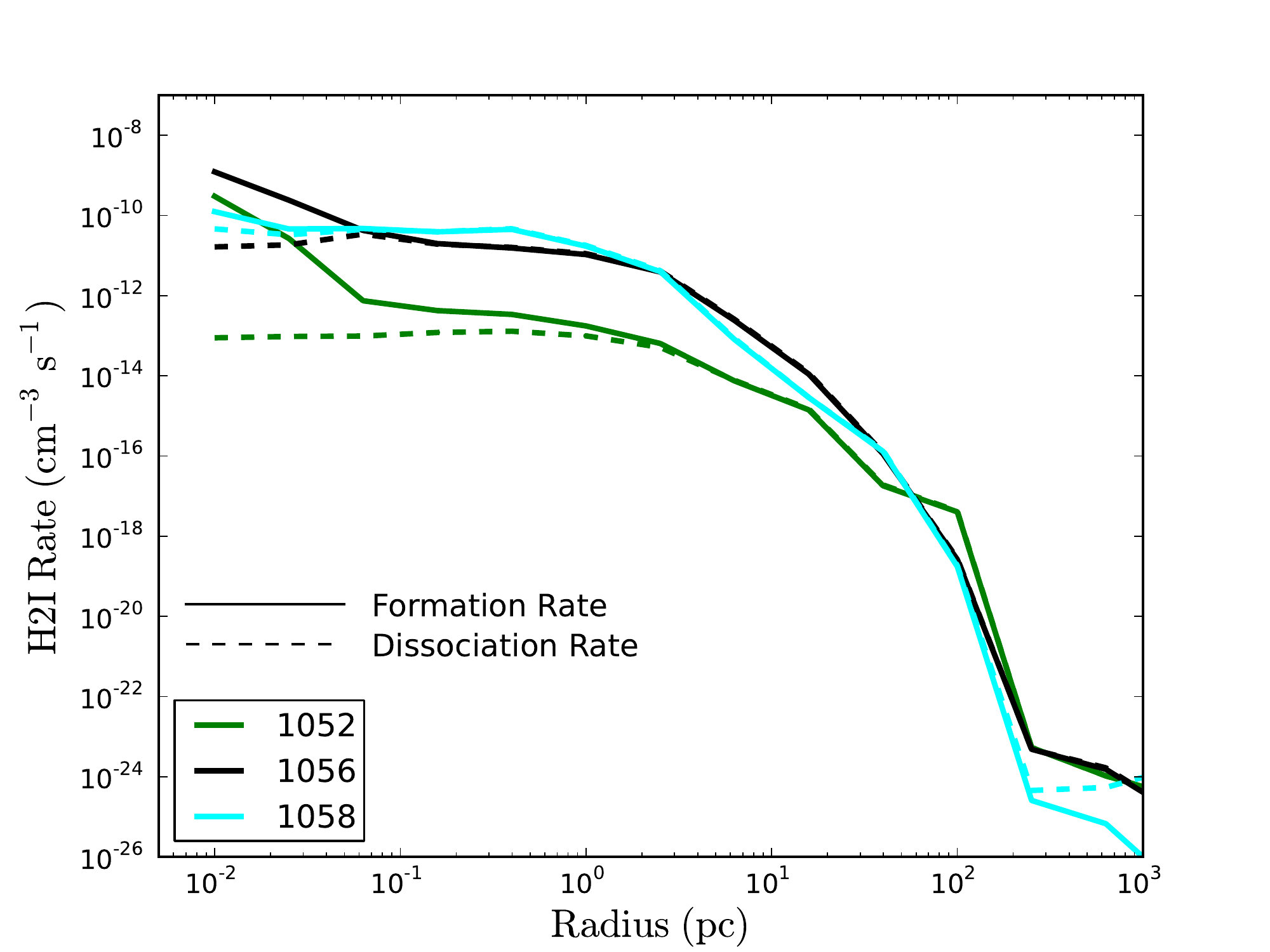}}
        \caption[]
        {\label{J21}
          The left hand panel shows the value of the flux, J, calculated along the 
          same 500 lines of sight from the source to the center of maximum density as was 
          done for the shielding factor calculation. The value of J varies from less than 
          $1 \times 10^{-1}\ J_{21}$ to greater than $1 \times 10^{6}\ J_{21}$ within 10 parsecs 
          of the center. J is calculated in this case at the point where the simulation 
          reaches the maximum refinement level. The right hand panel shows both the 
          \molH formation rate and the \molH dissociation rate computed using the same 500 
          sightlines and averaging over them. The solid lines are the formation rates while the 
          dashed lines are the dissociation rates. The \molH formation rate overwhelms the \molH 
          dissociation rate in the central regions. Only the results from simulations 1052,
          1056 and 1058 are shown for clarity. 
          }
      \end{center} \end{minipage}
  \end{figure*}


\section{Flux to Background Intensity Relation} \label{Sec:J21}
\noindent In all direct collapse simulations to date, in which a radiation module has been included, 
the authors have used a homogeneous and isotropic background UV flux to capture the effects of 
sources capable of dissociating \molHc. The radiation flux amplitude, $J$, is usually measured
in units of $J_{21} =\ $\J. This represents the intrinsic brightness or intensity
of a source which is assumed to be constant at all points in space.  The spectral 
flux at each point in space is then easily recovered by integrating over all angles. For the case of 
a stellar source the result of the integration is $\pi$ \citep[see e.g.][page 211]{Bradt_2008} 
while for the case of an isotropic field the result of the integration is 4$\pi$.\\
\indent In our simulations we neglect any contribution from background sources and 
calculate only the intensity from a single anisotropic close-by source of high intensity.
In Table \ref{Table:FluxTable} we show the flux of the source in each of our simulations. 
The source is placed at the same point in each simulation. Thus calculating the 
observed flux at the point of maximum density, when the source is switched on,
is straightforward. In this case $J$ is given by: 
\begin{equation} \label{J21eqn}
J = {\mathrm{Photons\ emitted\ per\ second} \times h  \over 4 \pi^2 r^2}
\end{equation} 
where h is Planck's constant and $r$ is the distance from the radiation source to the 
point of maximum density under the assumption that the medium is optically thin at 
all points, the second factor of $\pi$ in the denominator accounts for the solid angle.
Furthermore, we have used the frequency value at the center of the LW band 
only in the above conversion with the effect that the frequency, $\nu$, cancels out above and 
below the line.
The value of $J$ is shown in column 4 of Table \ref{Table:FluxTable} in units 
of $J_{21}$. The values in Table \ref{Table:FluxTable} are calculated at the time at which the 
source turns on. An effective stellar temperature of $T_{\rm eff} \sim 50000$ K is required 
to produce a spectrum which peaks in the LW bands. This effective temperature is typical of massive 
stars with masses in excess of $M_* \gtrsim 50$ \msolarc. 

\section{Results} \label{Sec:Results}
\noindent Table \ref{Table:SimDetails} shows the values of a range of physical quantities when the 
refinement level reaches the maximum refinement level allowed in our simulation, which is 18 in 
this case.  The maximum comoving spatial resolution reached in our simulations is 
$\sim 1.5 \times 10^{-2}\ \rm{h^{-1}\ pc}$, while the maximum proper resolution reached near
the end of each realization is $\sim 2.5 \times 10^{-3}\ \rm{h^{-1}\ pc}$.
The results show a variation in the time of collapse of the object due to the source 
flux amplitude. We begin by examining the quantities in each simulation that affect the ability 
of the gas to shield against the dissociating radiation, doing so allows us to determine what 
level of flux is required to first of all dissociate \molH and then to restrict its 
abundance. Throughout the following sections we refer to the \emph{core} of the simulation
as the region within 1 parsec of the point of maximum density and to the \emph{envelope} as 
the region surrounding the core extending to approximately 30 parsecs.

\begin{figure*}
  \begin{minipage}{175mm}      
    \begin{center}
     
      \begin{tabular}{cc}
        \includegraphics[width=9cm]{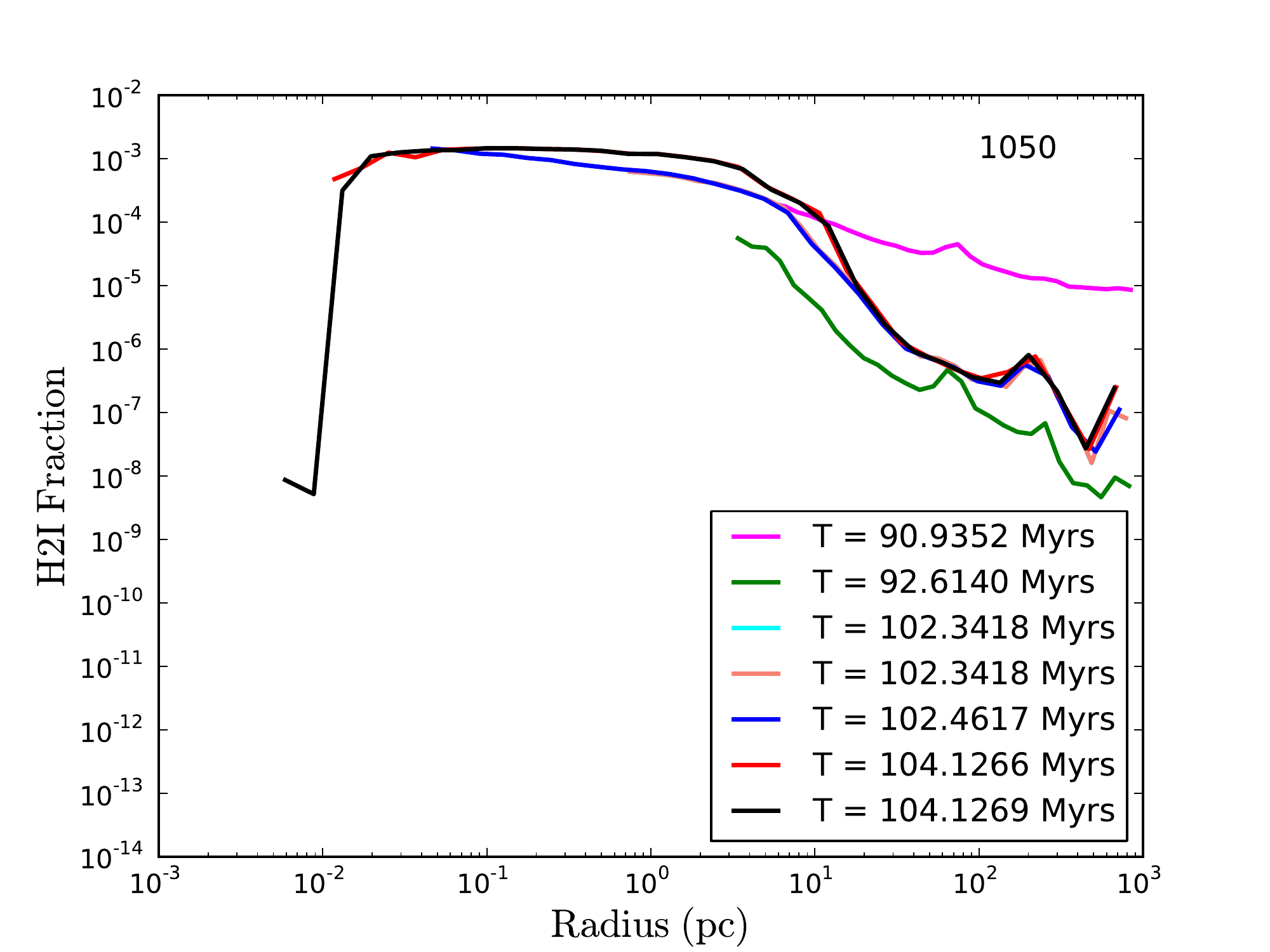} &
        \includegraphics[width=9cm]{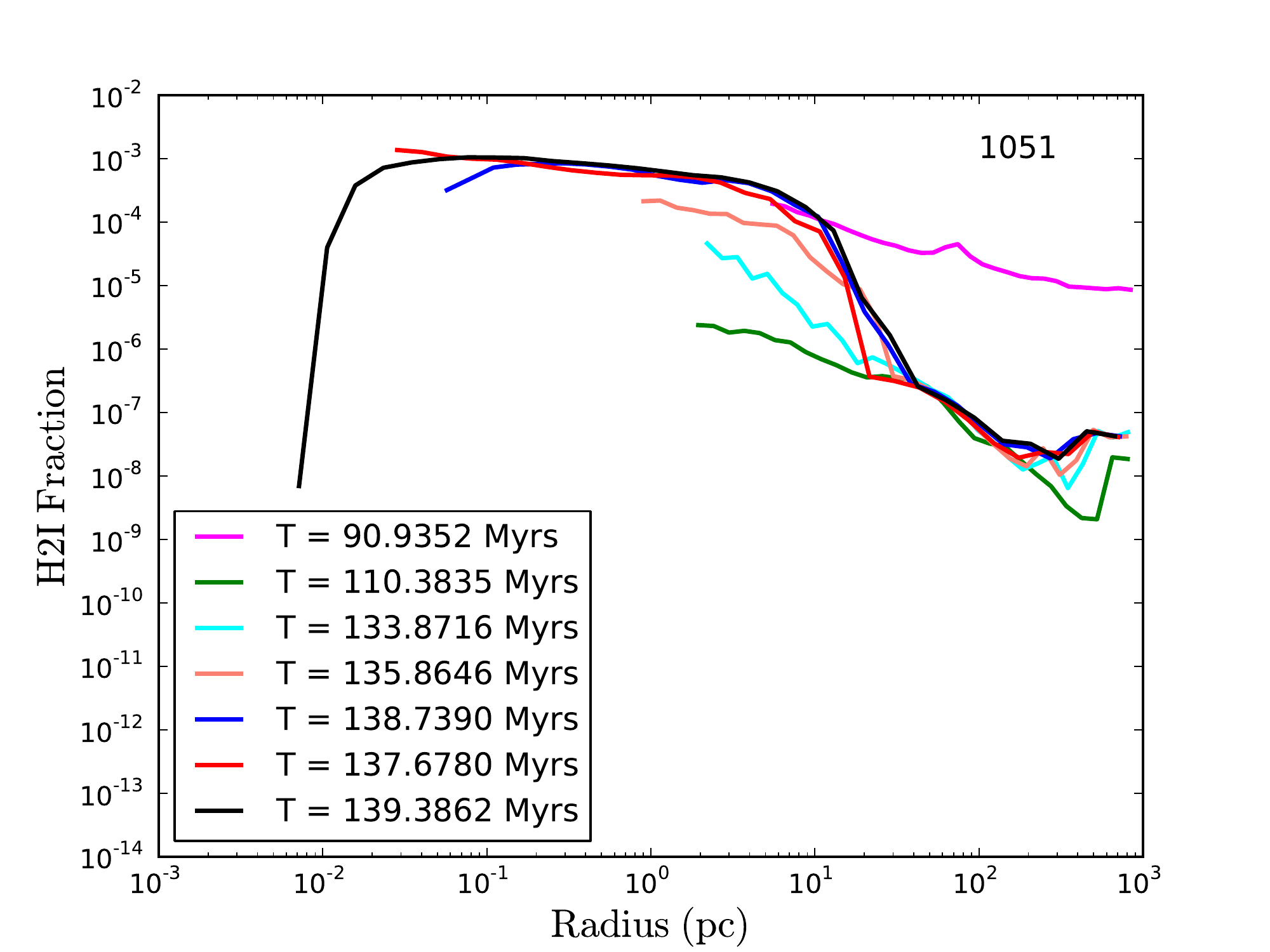} \\ 
        \includegraphics[width=9cm]{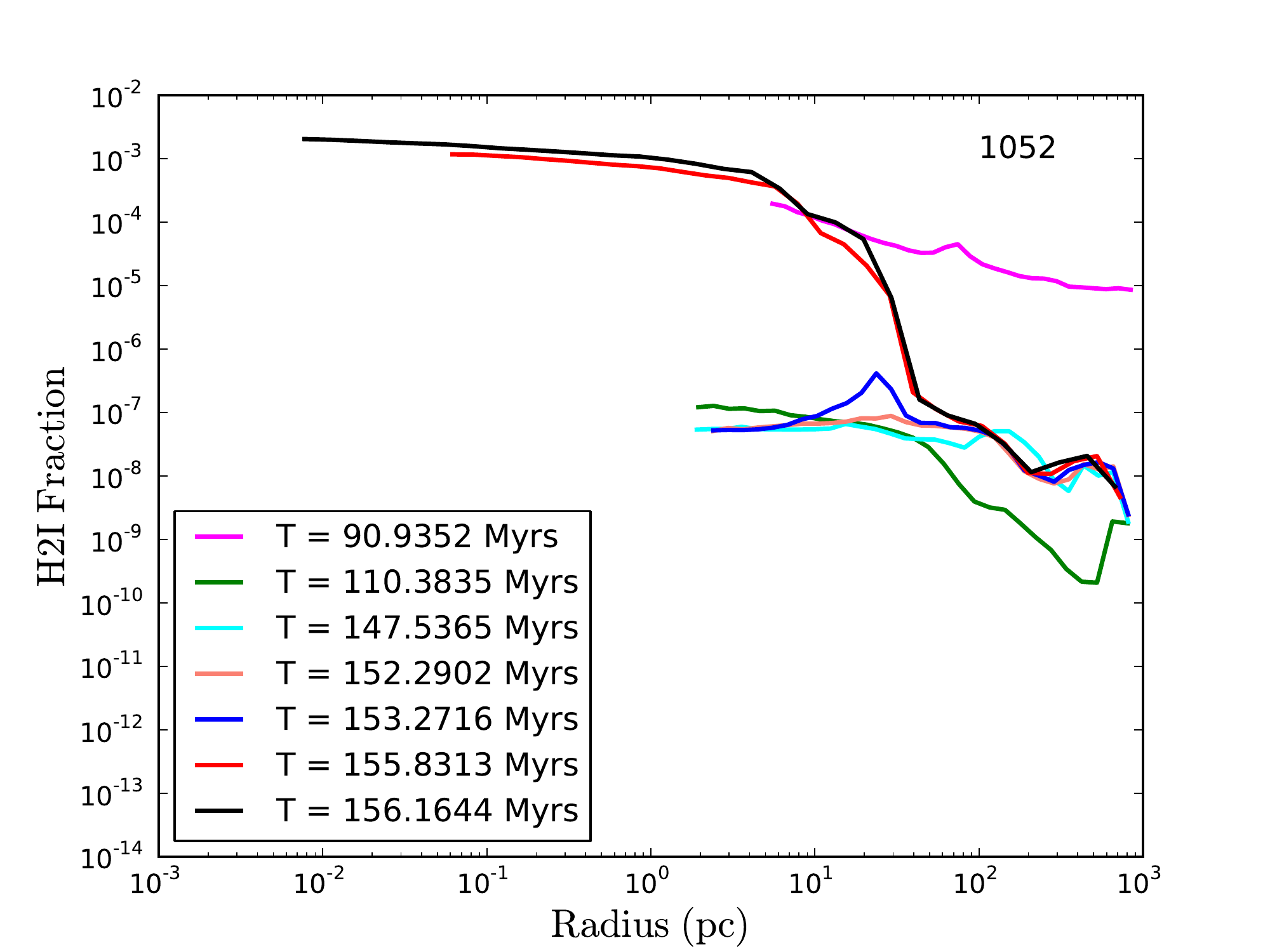}  &
        \includegraphics[width=9cm]{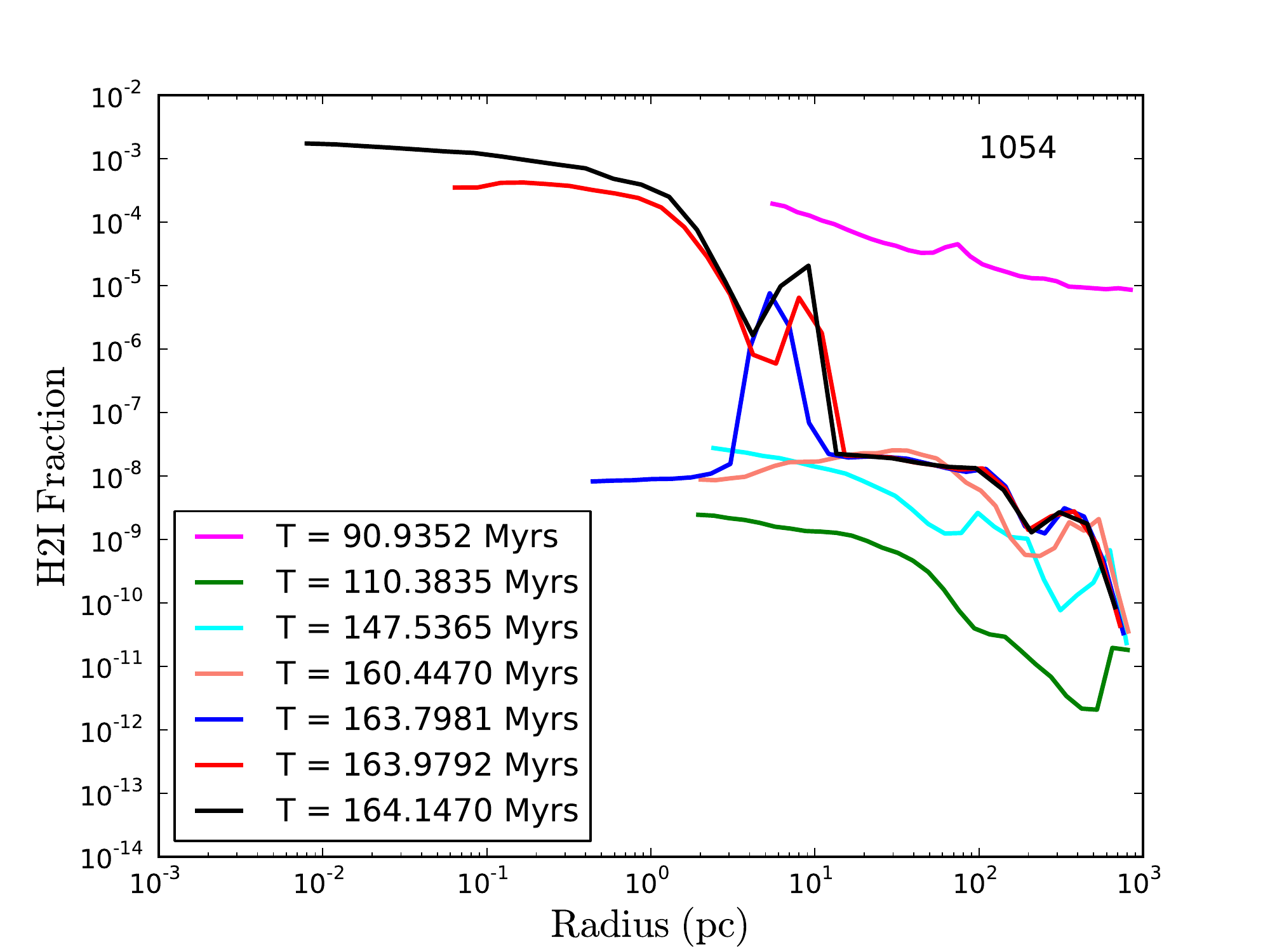} \\
        \includegraphics[width=9cm]{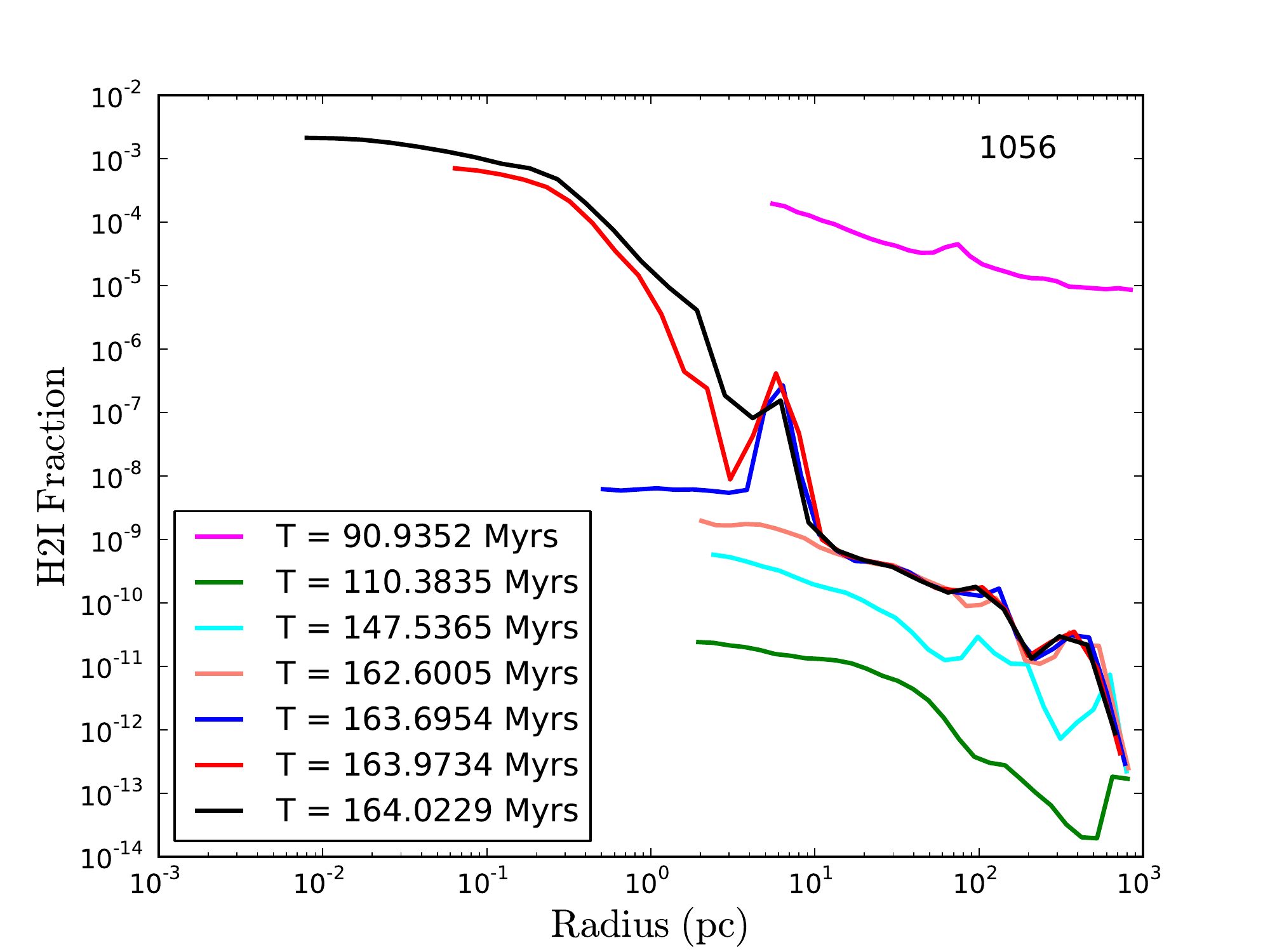} &
        \includegraphics[width=9cm]{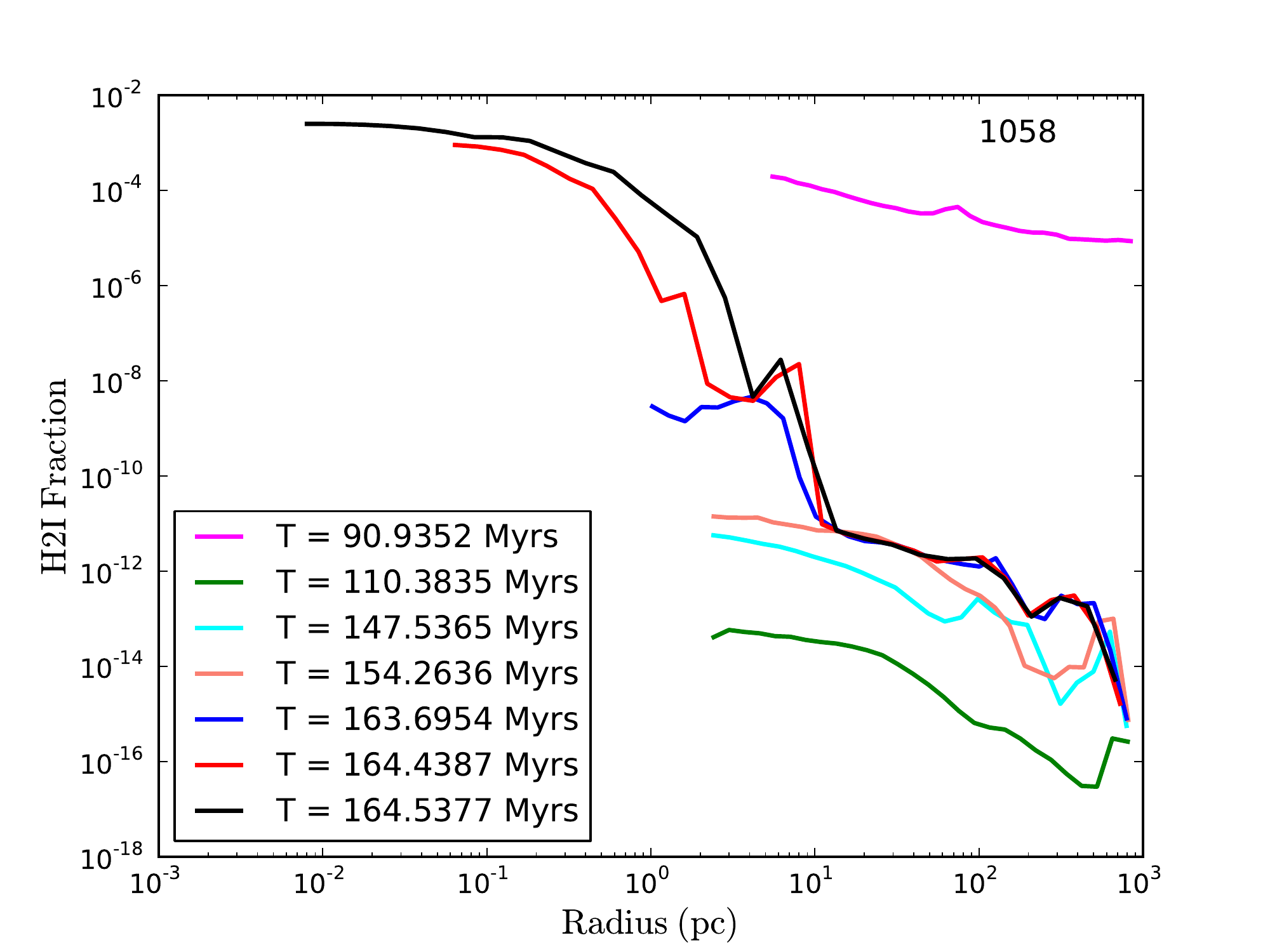}
      \end{tabular}
      \caption{ \label{H2ITimeSeries}
        Shown are the timeseries plots of the \molH fraction is each of 
        the simulations listed in Table \ref{Table:SimDetails}. The 
        source is switched on at 90.9352 Myrs (z = 32), at that  point the 
        \molH Fraction in the halo of interest is between $10^{-3}$ and $10^{-4}$.
        The \molH molecules in the halo are quickly dissociated and their density 
        reduced by several orders of magnitude. The \molH formation rate however 
        increases as the density of HI builds up and eventually overcomes the dissociation rate. 
        The halos in simulations 1050 and 1051 never reach atomic cooling status
        and they are always dominated by \molH cooling. In each of the other halos
        the fraction of \molH is suppressed sufficiently such that the halos cool
        predominantly via neutral hydrogen but note that \molH forms readily in the 
        core. 
    }
    \end{center} 
  \end{minipage}
\end{figure*}

\begin{figure*}
  \begin{minipage}{175mm}      
    \begin{center}
      \begin{tabular}{cc}
        \includegraphics[width=9cm]{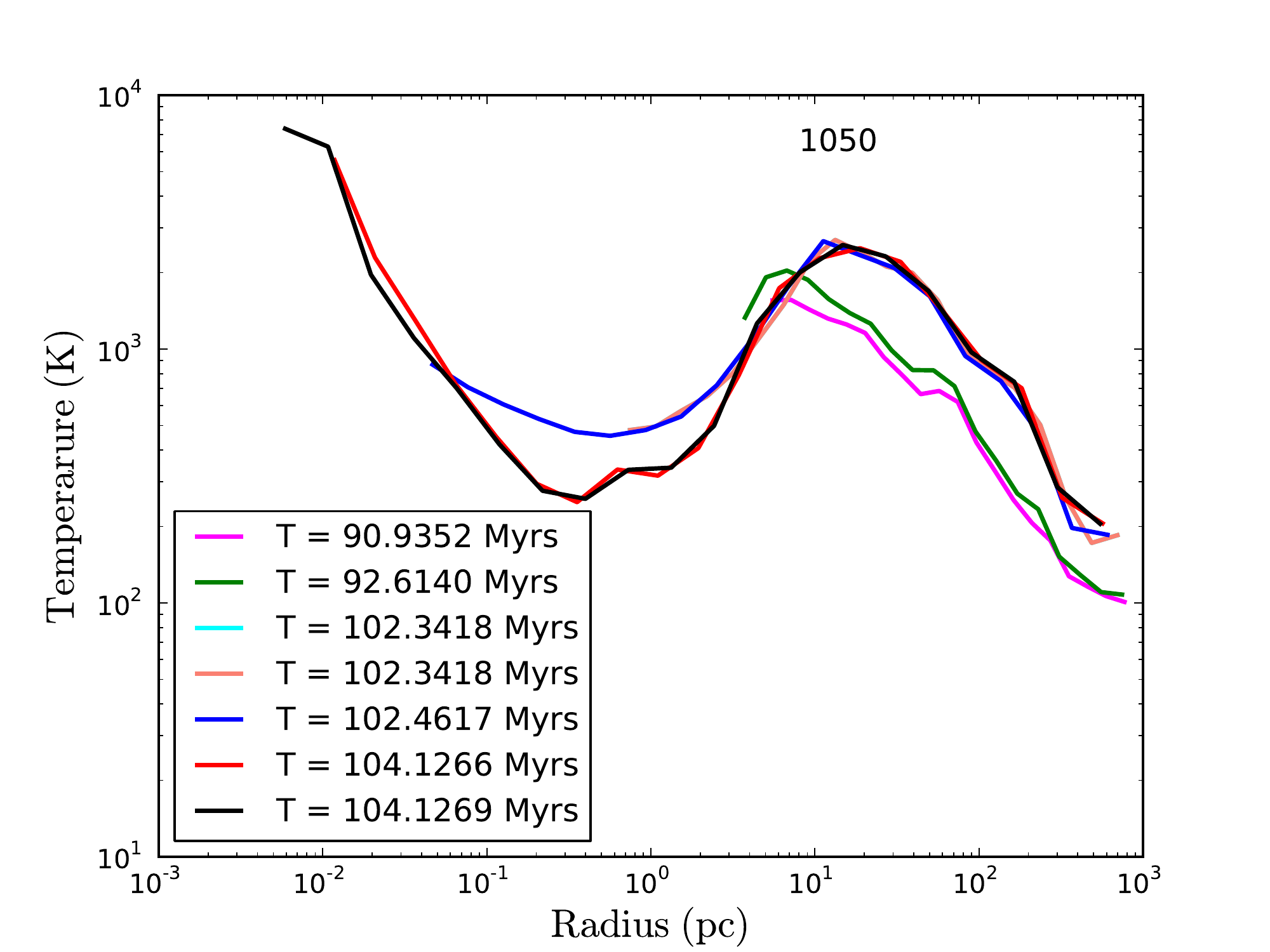} &  
        \includegraphics[width=9cm]{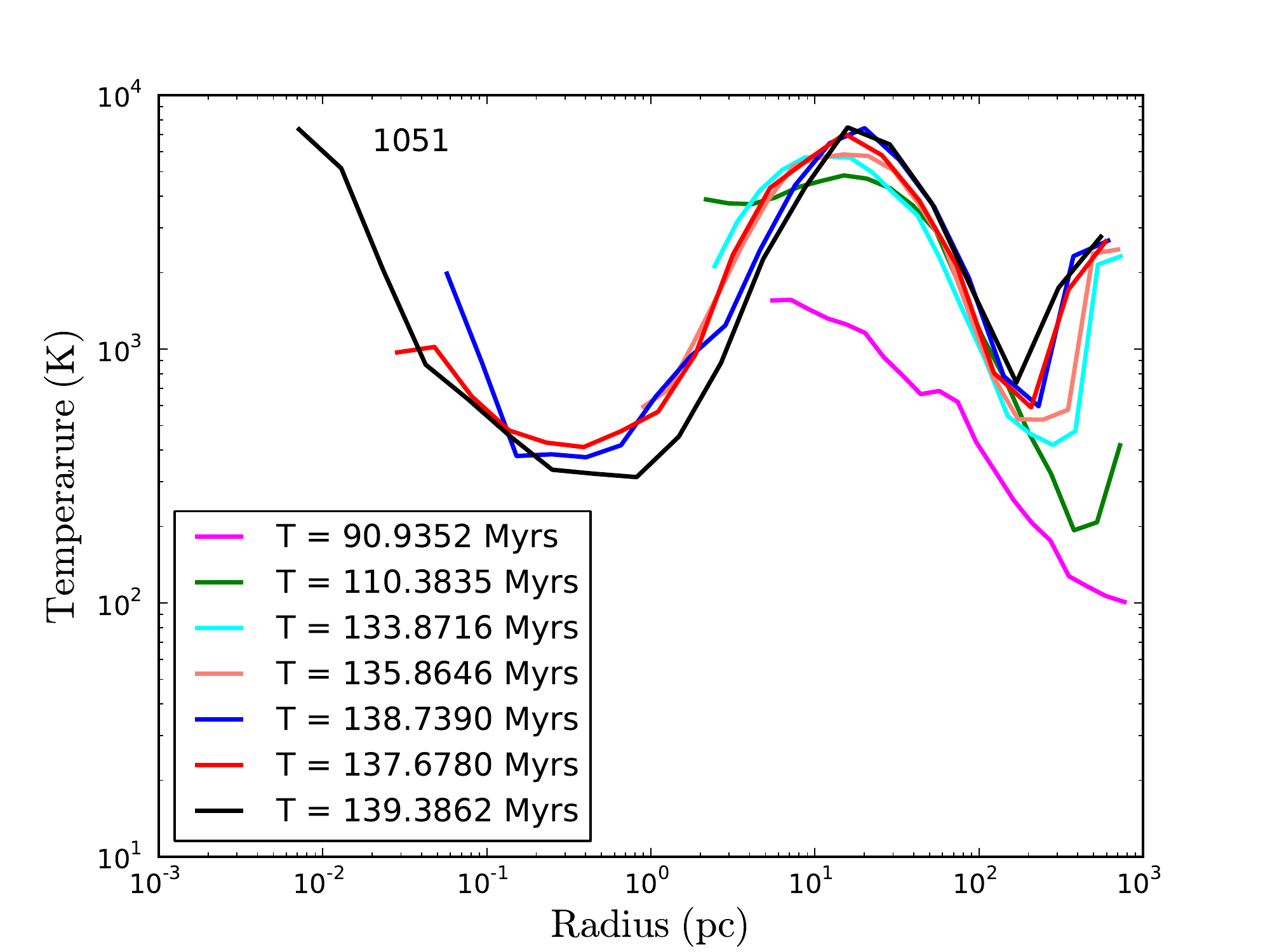} \\
        \includegraphics[width=9cm]{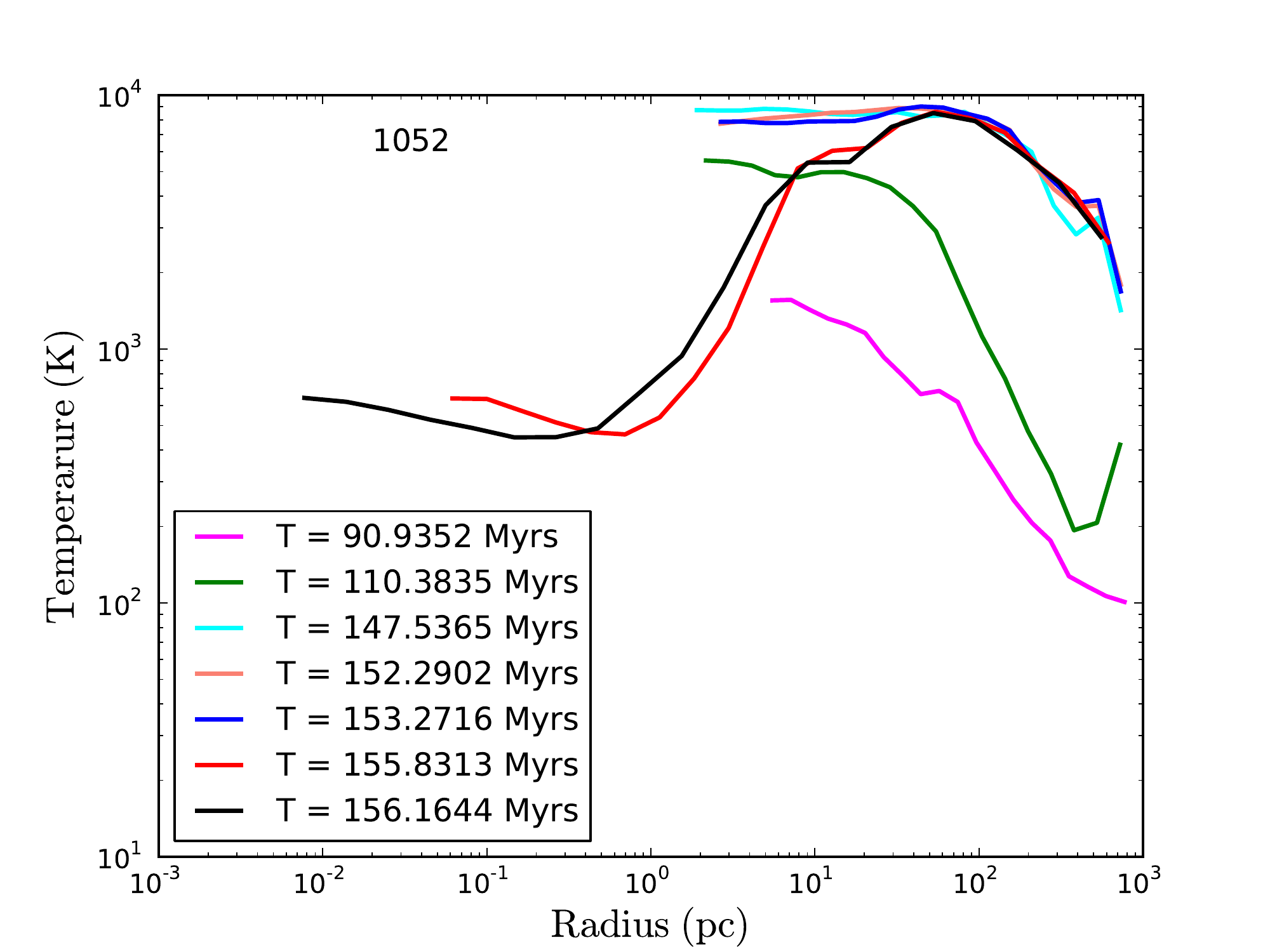} &
        \includegraphics[width=9cm]{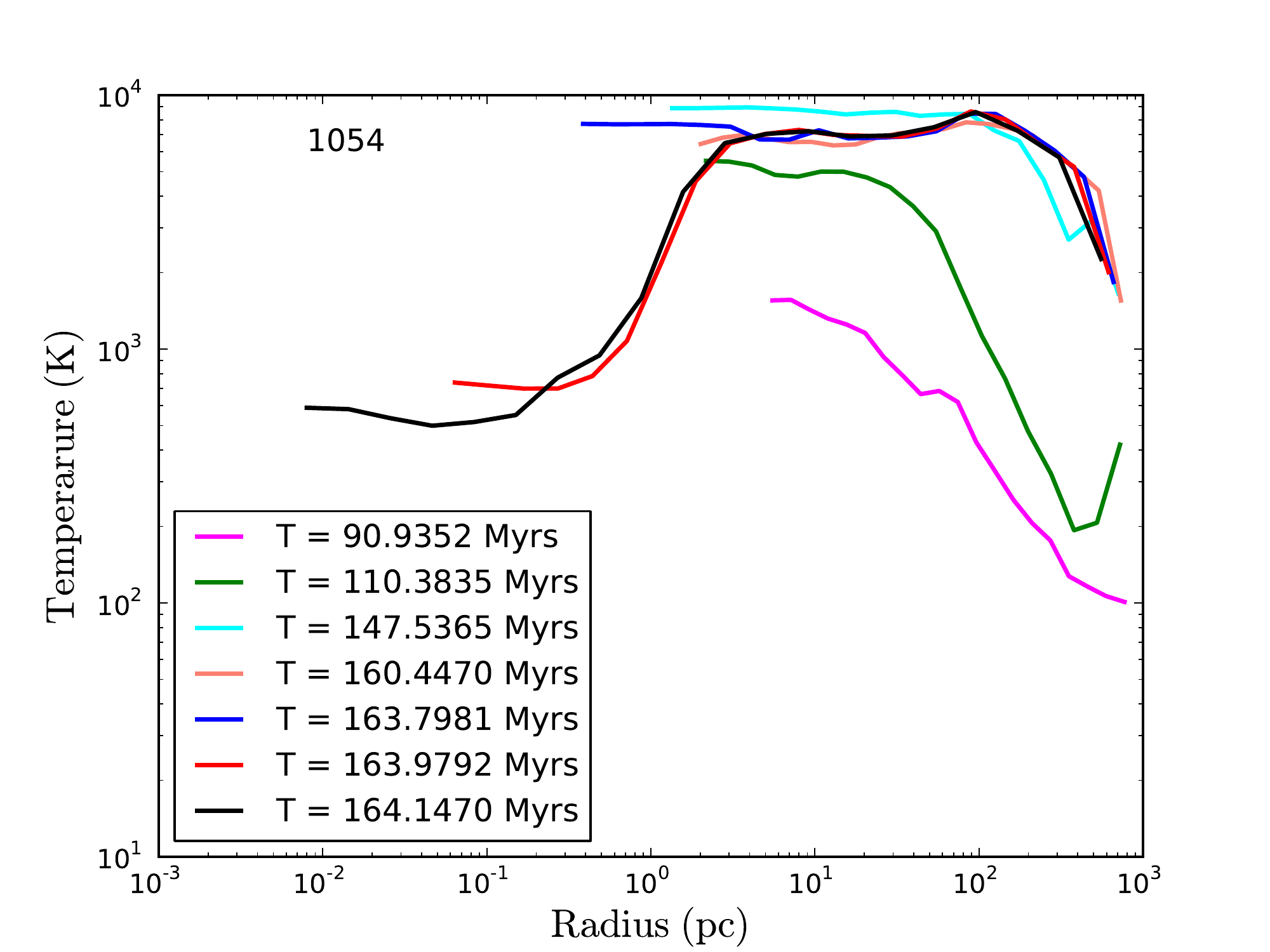} \\
        \includegraphics[width=9cm]{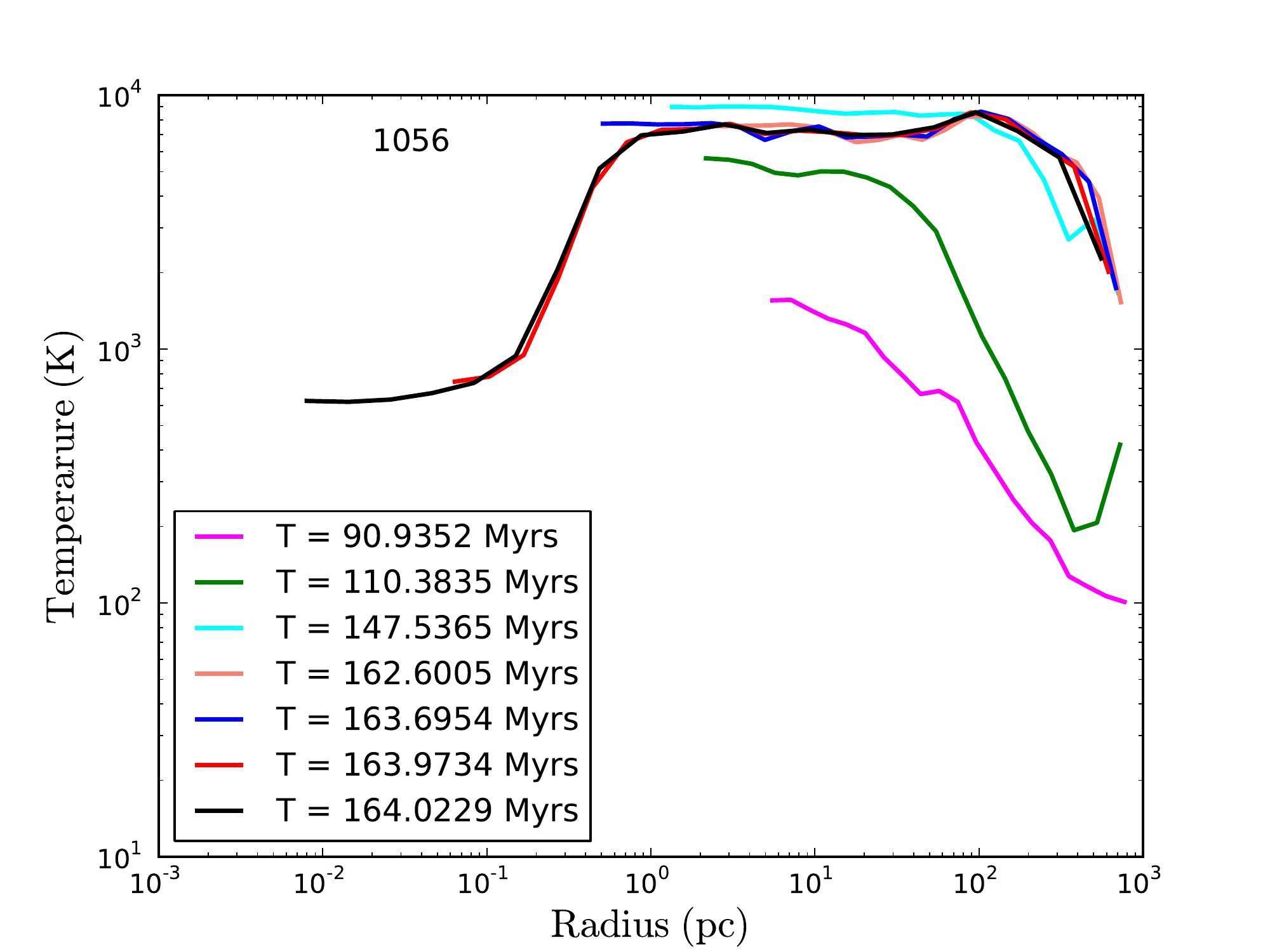} &
        \includegraphics[width=9cm]{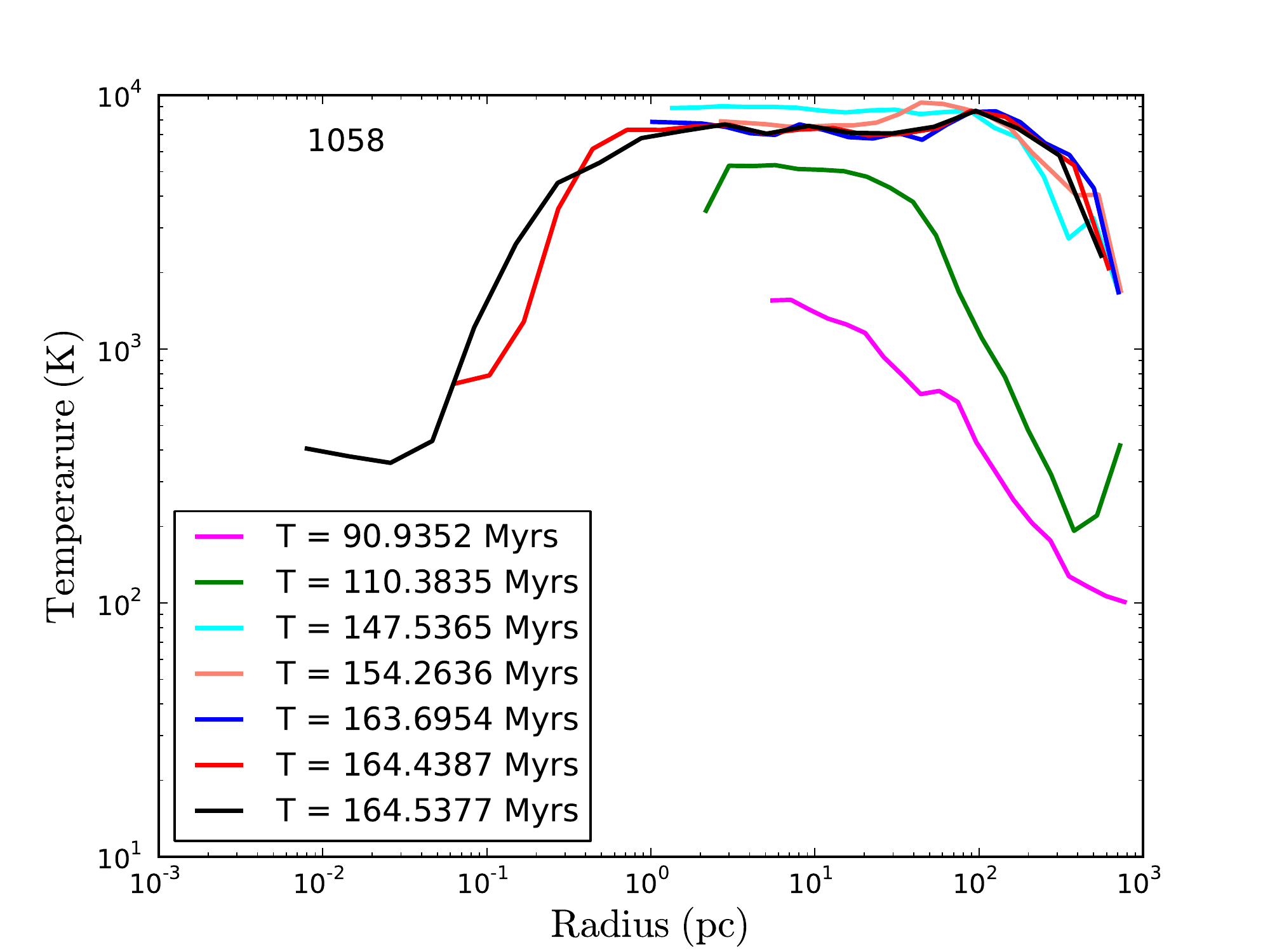}
      \end{tabular}
      \caption{ \label{TemperatureTimeSeries}
        The temperature timeseries for the same times and simulations as the 
        \molH fraction timeseries plots (Figure \ref{H2ITimeSeries}). The temperature is 
        initially close to $T\approx 10^3 \ \rm K$ with cooling by \molH prominent. The halo 
        in simulation 1050 is able to cool via \molH despite the dissociating source and 
        quickly collapses, the halo in 1051 is more resistant and must initially cool via HI
        (although its virial temperature is $T_{\rm vir} \ll 10^4 \ \rm K$). Nonetheless the halo 
        in 1051 is able to form \molH in large quantities once the HI density increases. This then
        triggers a collapse via \molH cooling. The other four halos experience stronger 
        dissociating fluxes and the temperature remains at $T \sim 10^4$K for a significant time. 
        Ultimately \molH is formed in the core in each case causing a sudden drop in the central 
        temperature. 
      }
    \end{center} 
  \end{minipage}
\end{figure*}


\begin{figure*}
  \begin{minipage}{175mm}      
    \begin{center}
      \begin{tabular}{cc}
        \includegraphics[width=9cm]{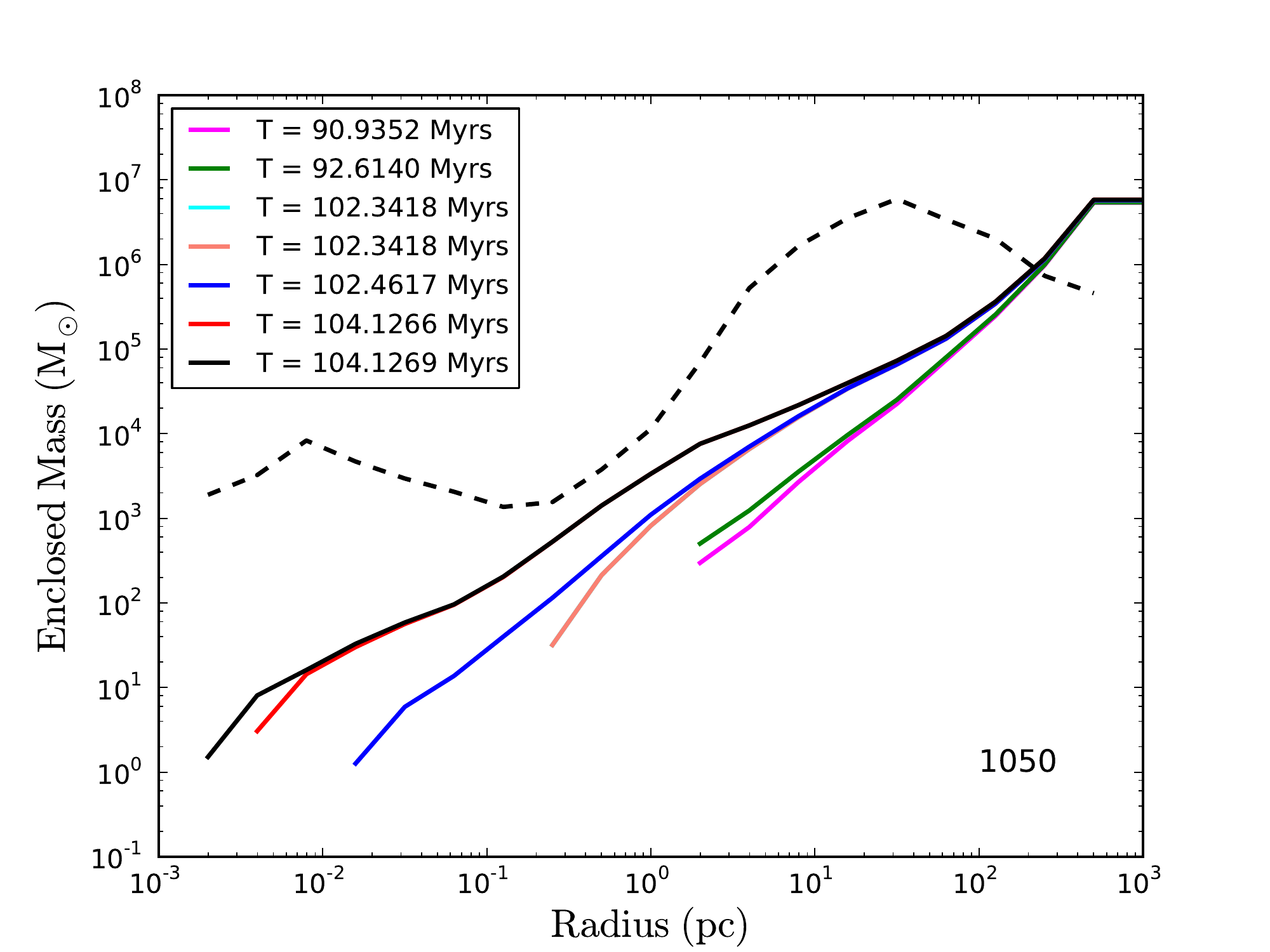}   &
        \includegraphics[width=9cm]{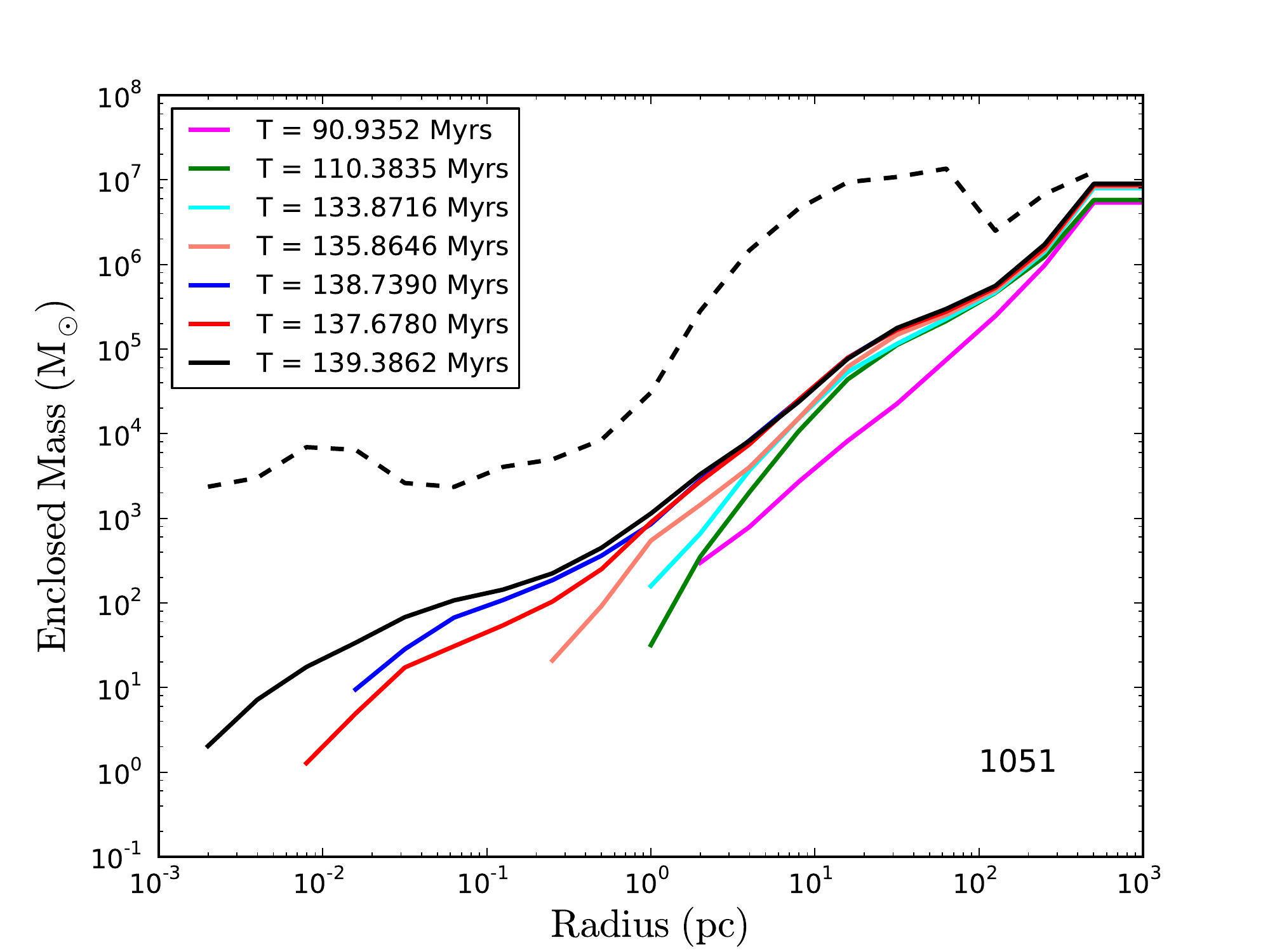}  \\
        \includegraphics[width=9cm]{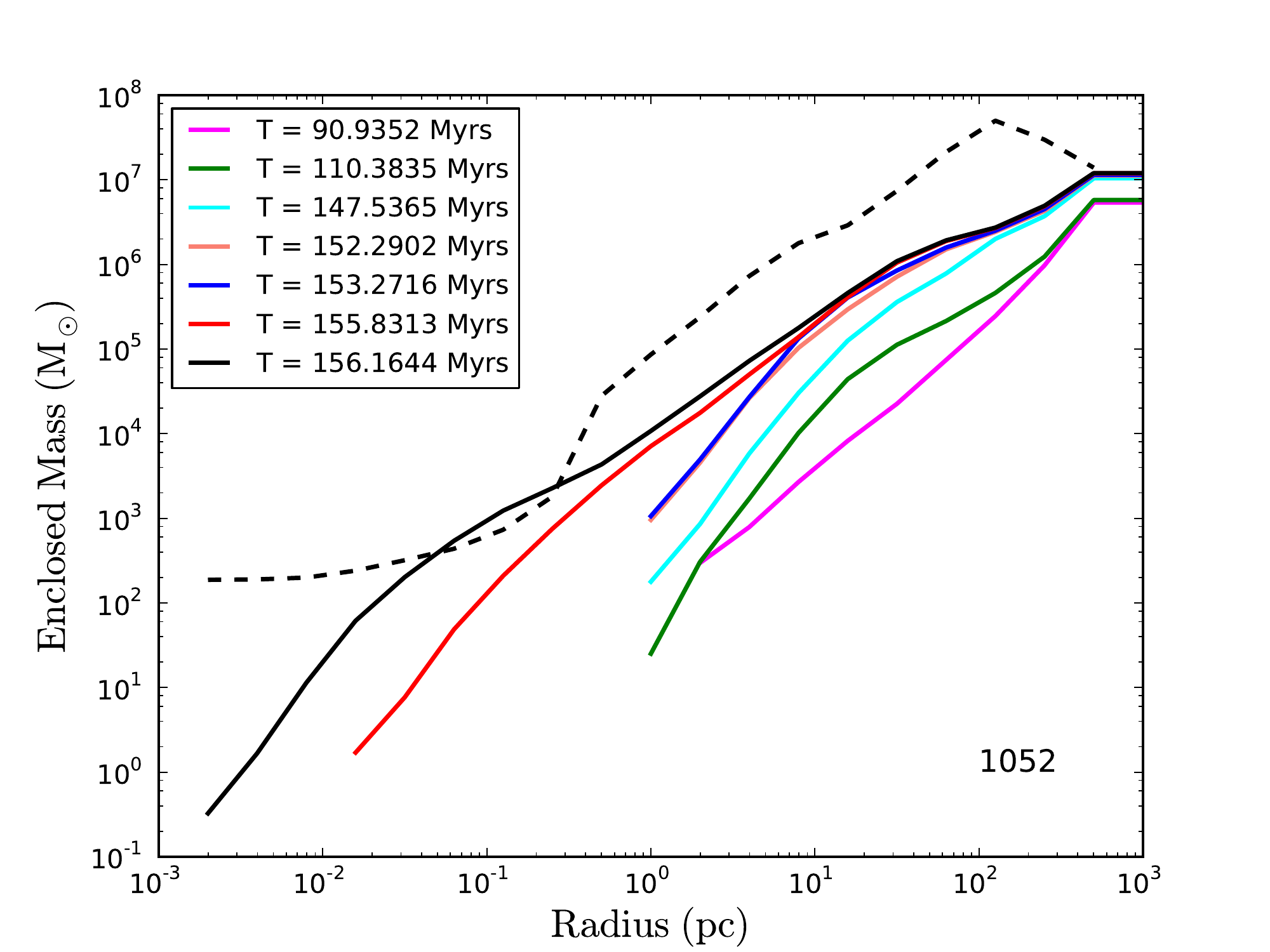}  &
        \includegraphics[width=9cm]{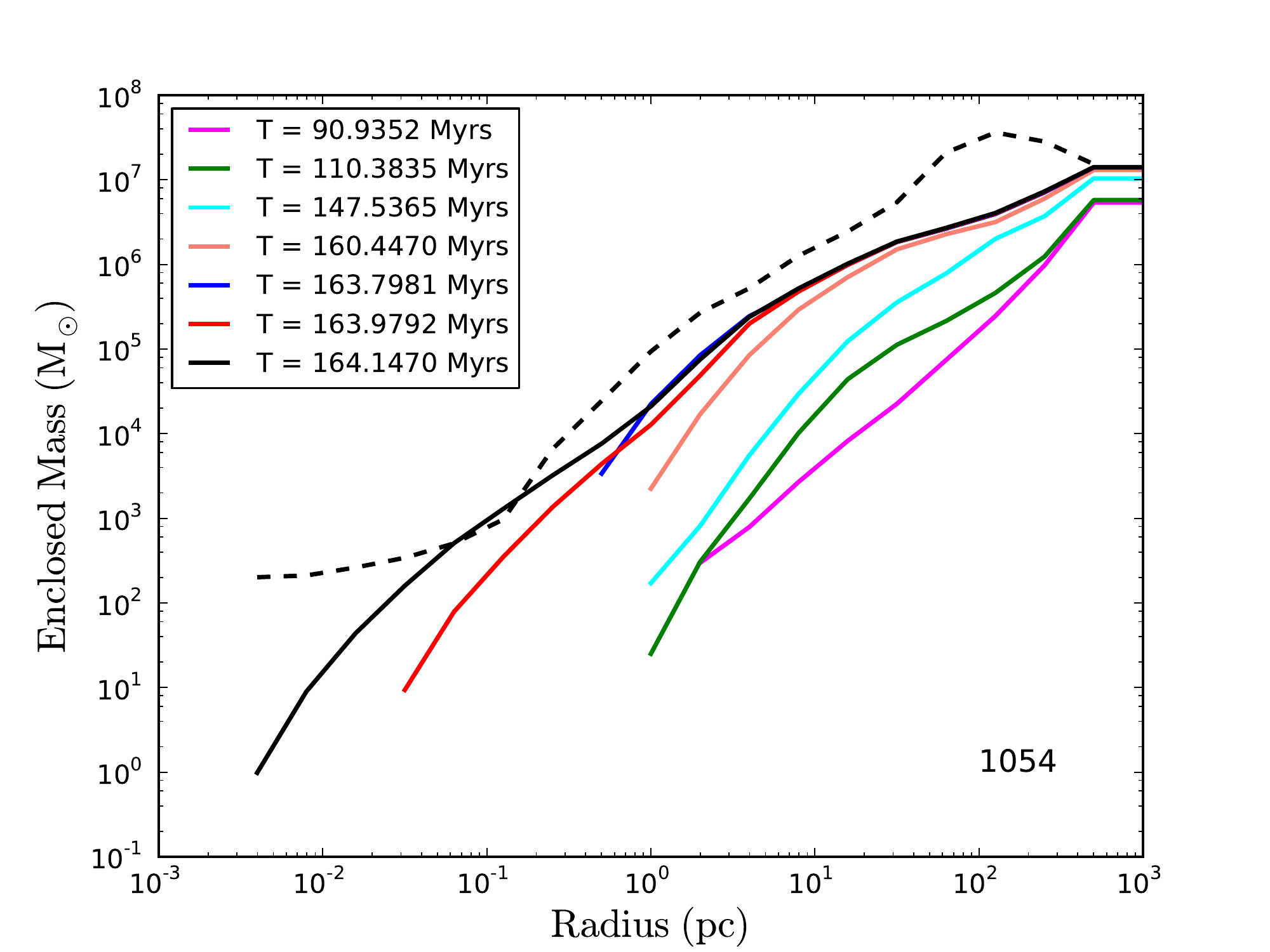}  \\
        \includegraphics[width=9cm]{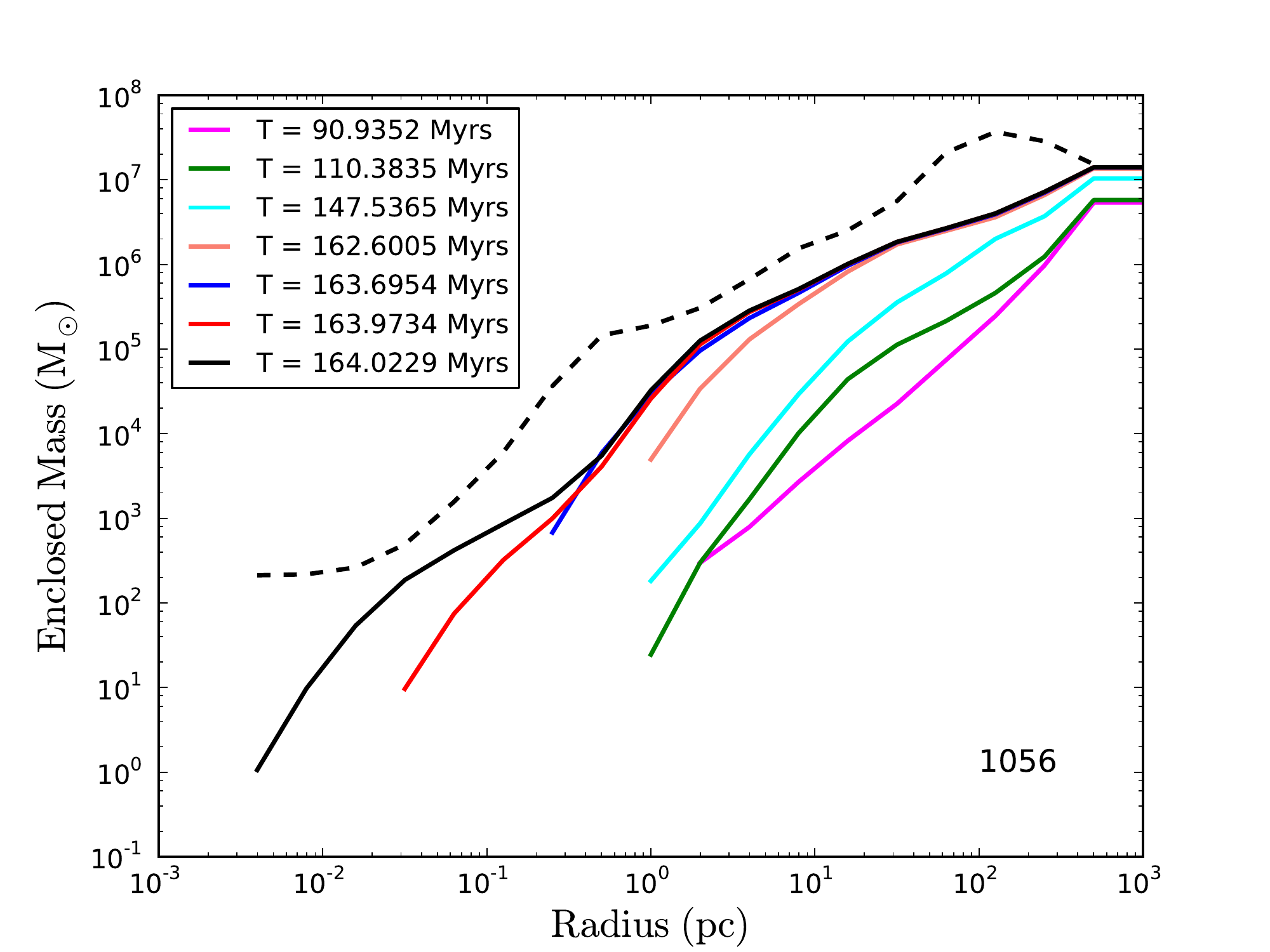}  &
        \includegraphics[width=9cm]{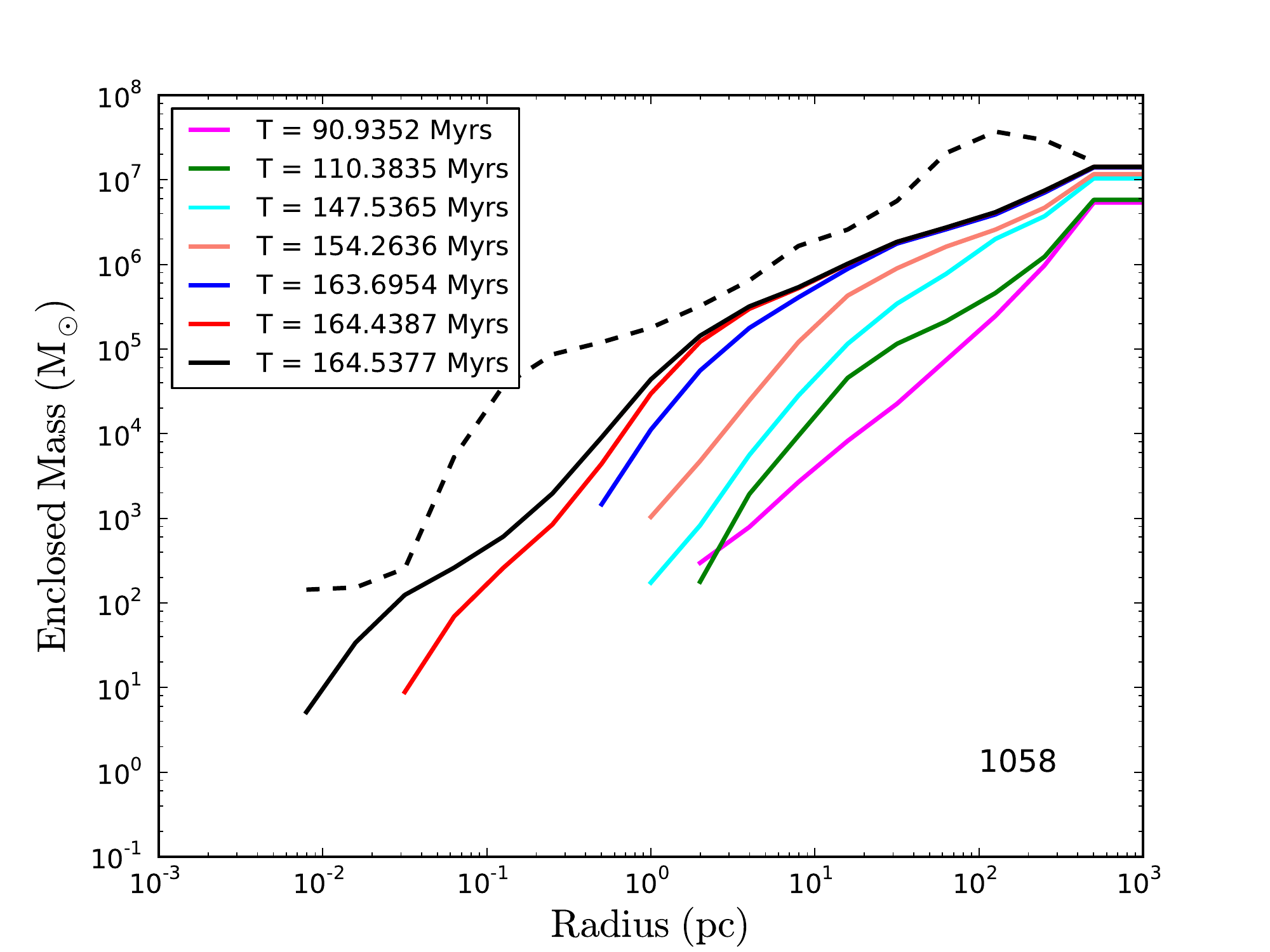}
      \end{tabular}
      \caption{
        \label{EnclosedMassTimeSeries}
        The enclosed mass for each simulation with the same output times 
        as Figure \ref{H2ITimeSeries} and Figure \ref{TemperatureTimeSeries}.
        The solid curves are the enclosed cell mass values while the 
        dashed lines give the Jeans mass at the final output time at the corresponding radius. 
      }
    \end{center} 
  \end{minipage}
\end{figure*}

\subsection{\molH Self-Shielding}
The left panel of Figure \ref{ShieldColumn} shows the shielding factor 
computed by averaging over 500 lines of sight between the source and the area surrounding 
the point of maximum density. All sight sightlines travel the same distance. 
In each simulation the shield factor outside of approximately 30 parsecs (outside the envelope) 
of the collapsing halo is 1.0, indicating that the medium is optically thin outside of this region. 
Within 30 parsecs the shielding factor quickly decreases to values between $10^{-5}$ and $10^{-6}$ 
approximately. There is also a trend for higher fluxes to enter the optically thick regime
at distances closer to the center of the maximum density as expected, e.g.
simulation 1058 only enters the optically thick regime at approximately a distance of 3 parsecs.
The shielding factor also displays significant scatter between simulations, this
is expected given the shielding factor (see equation \ref{DB37}) is a function
of both the column density and the temperature along a given sightline. \\
\indent The right hand panel of Figure \ref{ShieldColumn} shows the \molH column density computed
along the same 500 sightlines and also averaged. Similar to the shielding factor
there is a trend towards higher fluxes showing a smaller \molH column density as 
expected. The horizontal dashed line in the figure at $\rm{ 5 \times 10^{14}\ cm^{-2}}$
shows the point at which the medium is no longer assumed to be optically thin and 
where \molH self-shielding begins \citep{Draine_1996, Wolcott-Green_2011}. The maximum 
molecular hydrogen column density reached at the very center of the 
collapsing halo is between $\rm{1 \times 10^{19}\ cm^{-2}\ and\ 1 \times 10^{21}\ cm^{-2}}$.\\
\indent The left hand panel of Figure \ref{J21} shows the value of $J$ along the same 500 
lines of sight connecting the source and the point of highest density. The 
values are computed at the end of the simulation - as the simulation reaches the point of
maximum refinement. Within 100 parsecs the value of $J$ varies from $J \sim 10^{-2}\ J_{21}$ to 
$J \sim 10^{6}\ J_{21} $. For reference the global background at this redshift is expected 
to be $\rm{J_{LW\ Global} \lesssim 1.0\ \times \ }J_{21} $ \citep[e.g.][]{Dijkstra_2008}. \\
\indent Finally, the right 
hand panel of Figure \ref{J21} shows the formation rate and the dissociation rate of \molH for 
three selected simulations (1052, 1056 \& 1058) at the highest refinement level. The solid line 
is each plot represents the \molH formation rate while the dashed line of the same color 
represents the dissociation rate. The formation rate is calculated using the fitting formulae 
as used in the \enzo code \citep{Abel_1997} while the dissociation rate is calculated 
self-consistently by the radiative transfer module (see \S \ref{Sec:RadTransferSetup}).
For simulation 1052, we see up until approximately 100 parsecs from the center of maximum 
density that the dissociation rate matches perfectly the formation rate and no new \molH can be 
created (further from the halo the dissociation rate exceeds the formation rate). However, 
within 100 parsecs the formation rate for simulation 1052 overwhelms the dissociation rate 
and \molH can form. A similar characteristic is shown for both simulations 1056 and 1058. 
However, in the case of 1056 and 1058 the formation rate is only able to exceed the dissociation 
rate much closer to the center of the halo, at approximately 5 parsec and 2 parsec distances, 
respectively. In all cases this indicates that \molH is readily formed in the core of the halo, 
even in the presence of extremely strong fluxes, due to the increase in the \molH formation rate 
compared to the dissociation rate.\\

\subsection{Physical Characteristics of the Illuminated Halo}

\begin{figure*}
 \centering 
  \begin{minipage}{175mm}      \begin{center}
      \centerline{\includegraphics[width=14cm]{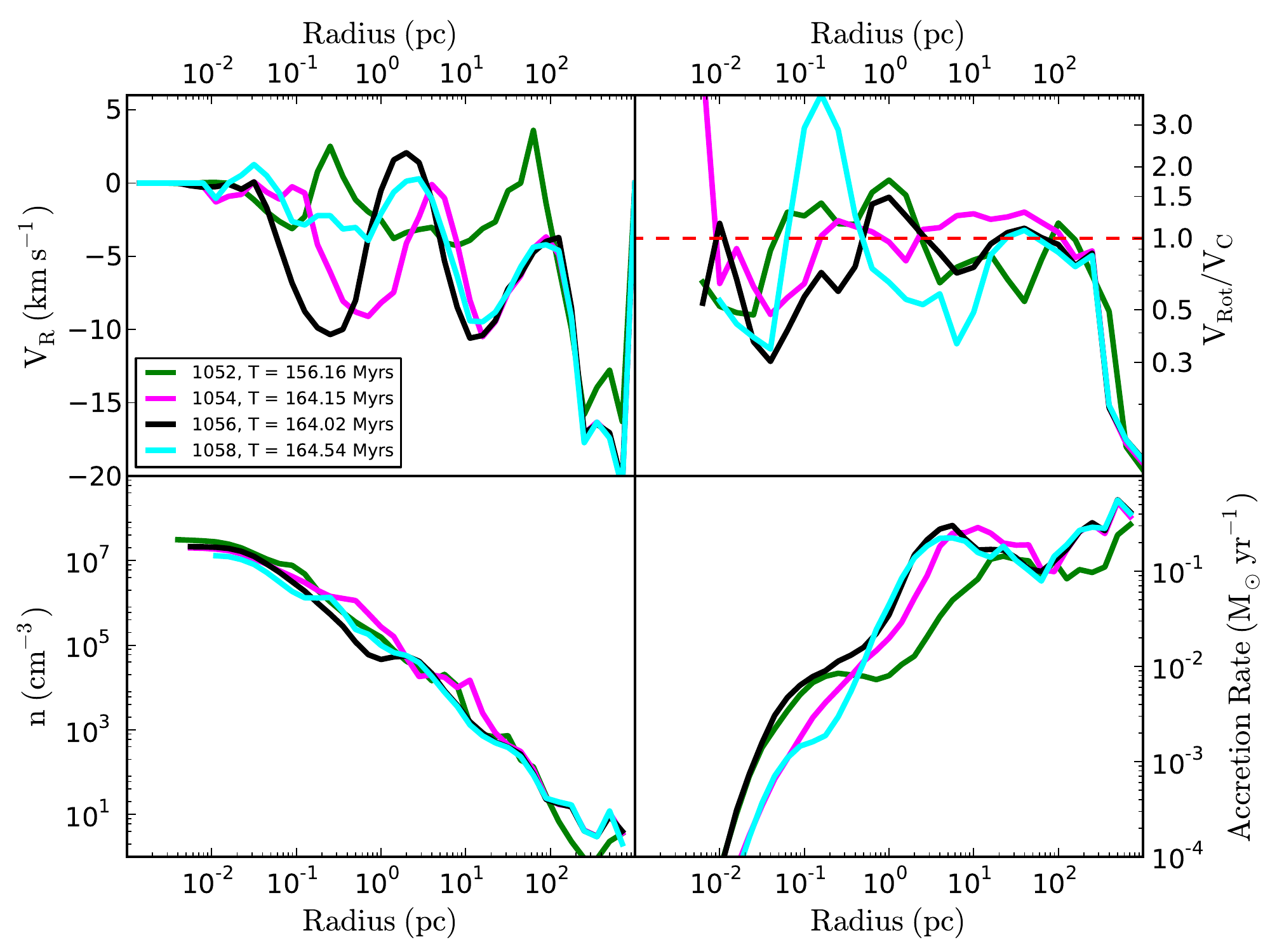}}
    \caption[]
    {\label{FourPanelPlot}
      All halos that reach atomic cooling status are shown. 
      The plots are at the time each simulation reaches the maximum refinement level. 
      The bottom left panel shows the gas number density, $n$, plotted as a 
      function of radius. The top left panel shows the radial velocity, $V_R$, as
      a function of radius. The top right panel shows the ratio of the 
      rotational velocity to the circular velocity, $V_{Rot}/V_C$, as a function of radius.
      The bottom right panel shows the mass accretion rate, $\dot{M}$, as a function of
      radius at the final output time. }
  \end{center} \end{minipage}
\end{figure*}


\begin{figure*}
  \begin{minipage}{175mm}      
    \begin{center}
      \begin{tabular}{cc}
        \includegraphics[width=15cm]{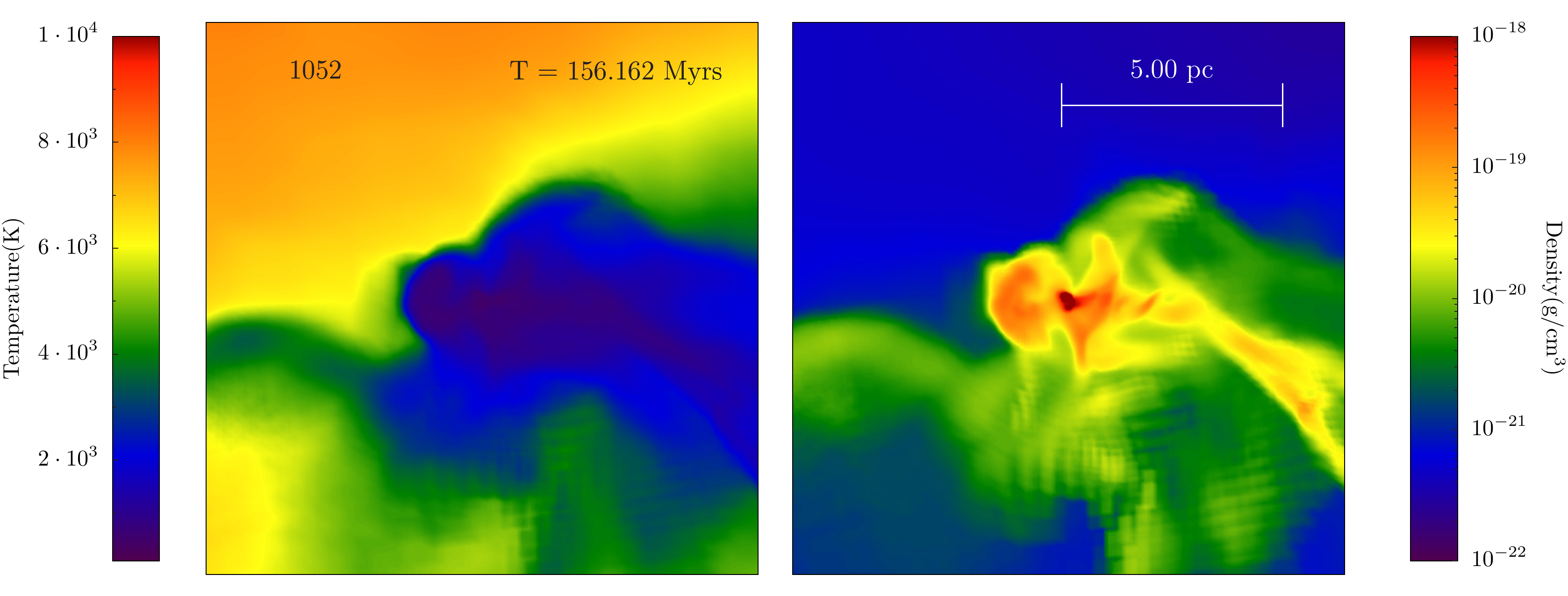}  \\
        \includegraphics[width=15cm]{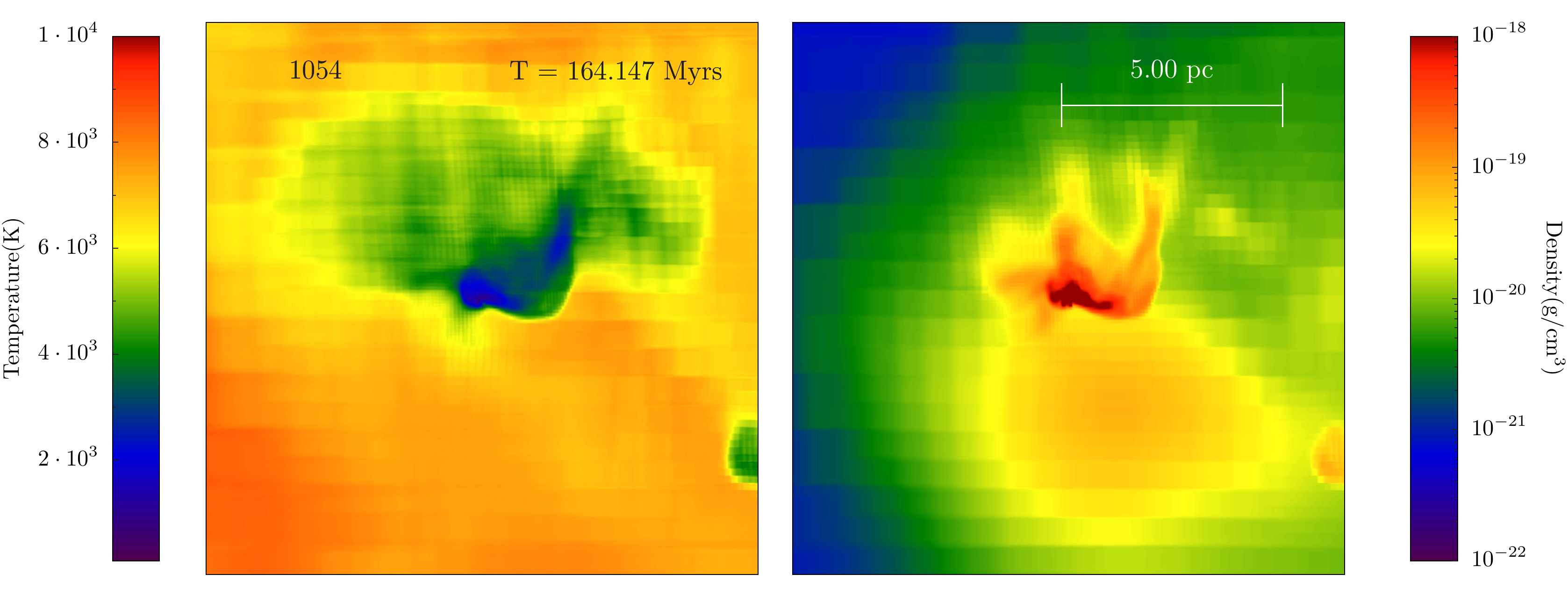}  \\
        \includegraphics[width=15cm]{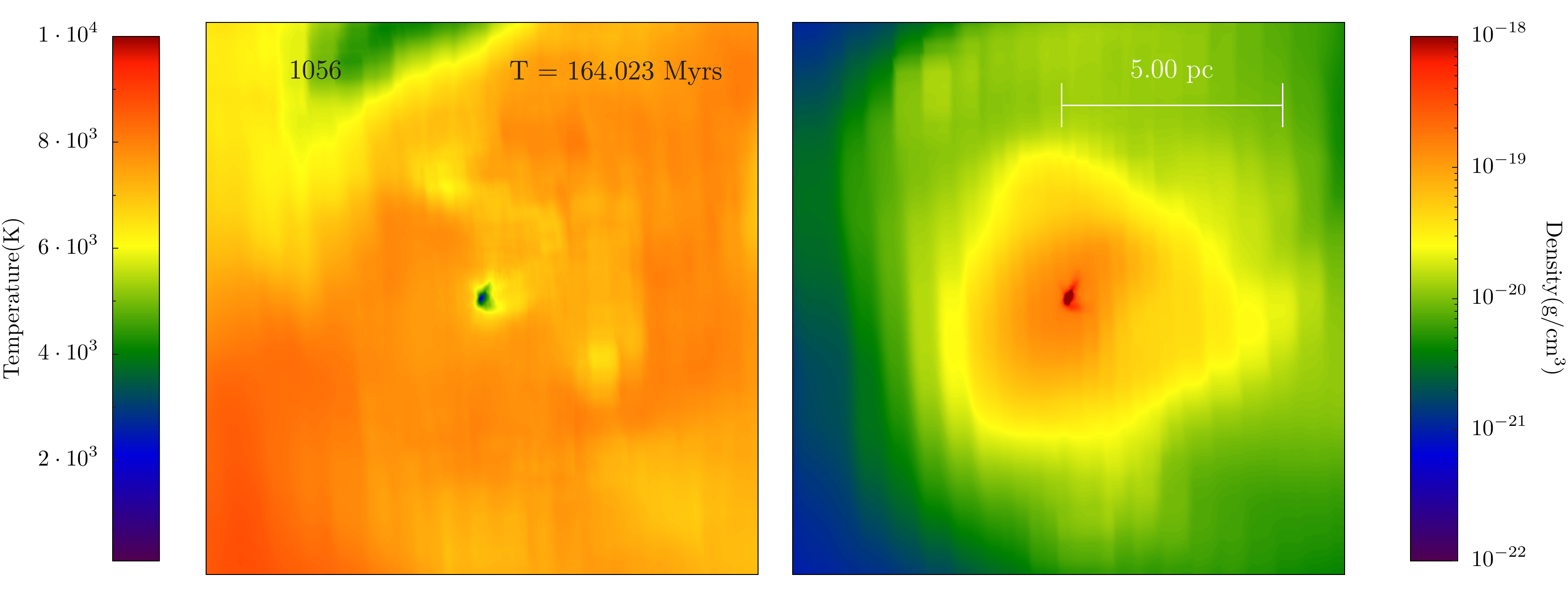}  \\
        \includegraphics[width=15cm]{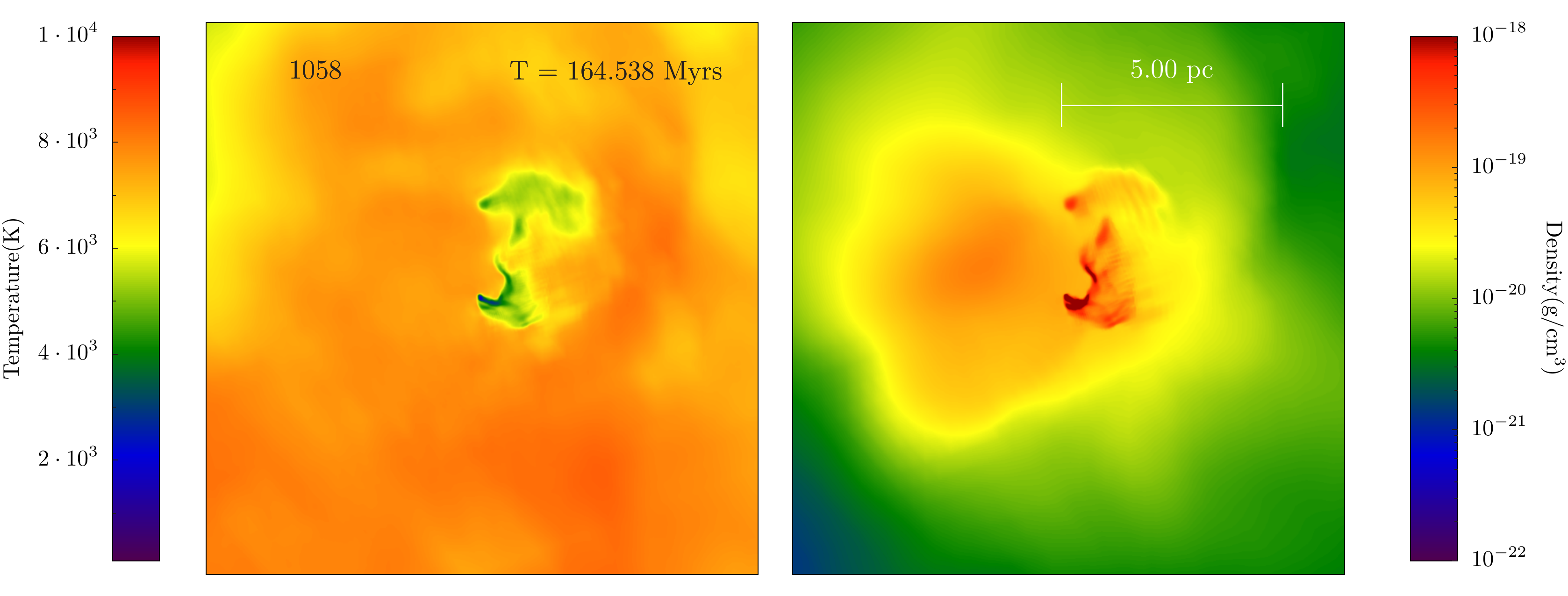}  \\
      \end{tabular}
      
      \caption{\label{TempDenProjs}
       Each halo is shown in projection as it reaches the highest refinement level.
       The projection in each case is made along the angular momentum 
       vector. The left panel of each row shows the temperature across approximately 
       10 parsecs. The right panel in each row shows the density in the same region.
       The projections are shown only for the fours halos which reach the atomic 
       cooling threshold. The spatial scales of all panels are identical.
    }
    \end{center} 
  \end{minipage}
\end{figure*}


\subsubsection{Time Series Analysis}
\noindent We begin by looking at a time series analysis of the collapsing halos when they are 
subjected to different \molH dissociating fluxes. In Figure \ref{H2ITimeSeries} we show the 
\molH fraction for each of our simulations, where each panel is for a different source flux. 
We use radial profiling to examine the quantities. Radial profiling 
allows us to best determine the properties of the gas surrounding the core and envelope. 
While the radiation is anisotropic and thus comes from a preferred direction, the dominant 
gravitational forces acting during the collapse do not have a preferred direction and so in 
this case radial profiling best captures the state of the gas for all but the very 
earliest stages after the source is switched on. \\
\indent In each case we start at the time at which the source was switched on (T = 90.9352 Myrs, 
corresponding to $z=32$). Each simulation was run until the maximum refinement level was 
reached. What is clearly noticeable from each simulation is that initially the \molH 
fraction decreases by a large factor, with larger drops for larger flux amplitudes as 
expected. For simulations 1050 and 1051 the decrease in the \molH fraction is quickly 
recovered. Simulation 1050 in particular reaches \molH values comparable to 
its initial value within 10 Myrs and subsequently collapses to form a minihalo - capable of 
forming population III stars \citep[e.g.][]{Abel_2002}. Similarly simulation 1051 displays 
a similar trend, albeit the collapse to a minihalo takes longer and is significantly delayed.
Both halos also display a strong decrease in \molH at the very center of the halo, as has been 
seen before in simulations of population III star formation \citep[e.g.][]{Turk_2012}. This is 
caused by the collisional dissociation of \molH as gas rapidly flows to the center of the 
newly formed potential. \\
\indent Simulations 1052, 1054, 1056 and 1058 all experience a strong dissociating flux. As a 
result the \molH fraction is strongly suppressed initially. For each of these fluxes the collapse 
takes place approximately 60 Myrs after the source switched on, equivalent 
to a delay of approximately 50 Myrs from the case it which no flux exists. Within the central 
10 - 40 parsecs the fraction of \molH increases rapidly due to the formation rate of \molH 
exceeding its dissociation rate (see Figure \ref{ShieldColumn} and Figure \ref{J21}). By the 
end of each simulation the \molH fraction in the core of the halo (within $\sim 1$ parsec) has 
returned to its equilibrium value of $1 \times 10^{-3}$. \\
\indent In Figure \ref{TemperatureTimeSeries} we plot the temperature profile for each of our 
simulations. Simulation 1050 is almost completely unaffected by the dissociating flux, the 
temperature remains close to $T = 10^{3}\ \rm K$ for most of the simulation. As the HI density 
grows the formation of \molH is enhanced at around 10 parsec distances and the temperature 
decreases to $T \sim 500 \ \rm K$. At the very end of the simulation and deep within the core 
($\rm{R \lesssim 1}$ parsec) where \molH is collisionally dissociated we do see the temperature 
significantly exceeding $T = 10^{3}\ \rm K$. Simulation 1051 behaves similarly, although the 
stronger flux, compared to simulation 1050, means that the temperature in the halo increases to 
closer to $T = 10^{4}\ \rm K$, in fact reaching values around $T \sim 8000\ \rm K$ at 
approximately 10 parsecs. At this point however, the rapid formation of \molH drives the 
temperature back down to closer to $T \sim 500\ \rm K$ similar to the 1050 case. \\
\indent The atomic halos (1052, 1054, 1056, 1058) instead show a rather different behavior. 
The initial strong suppression of \molH means that the temperature quickly rises. Within 
approximately 20 - 30 Myrs the temperature of the gas has reached $T\sim 10^4\ \rm K$. The main 
coolant is now HI and the virial temperature of the halo quickly exceeds $T\gtrsim 10^4\ \rm K$. 
The gas remains at $T_{\rm{Vir}} \sim 10^4\ \rm K$ as the halo begins its collapse. However, within $\sim$ 
10 parsecs of the center, the \molH fraction (see Figure \ref{H2ITimeSeries}) is able to 
grow considerably. The presence of \molH has a dramatic effect on the gas temperature enabling 
cooling back down to $T\sim 1000\ \rm K$. This is the case in the centers of each of the 
atomic halo simulations. \\
\indent Finally, in Figure \ref{EnclosedMassTimeSeries} we show the enclosed gas mass as a 
function of radius from the point of maximum density. The solid line in each panel is the enclosed 
mass. For reference the Jeans Mass, at that radius, is plotted (dashed black line) for the final 
output time. Once the Jeans mass is exceeded the gas becomes gravitationally unstable to collapse. 
Both simulations 1050 and 1051 fail to exceed the Jeans threshold at any radius (although 1050 
does show signs of instability as expected close to $M \sim 10^3 $ \msolarc). In both cases the 
gas is collapsing via \molH cooling and has yet to form a self-gravitating clump by the end of 
the simulation. However, both are expected to collapse within a short time as the clump accretes 
mass and the enclosed mass increases. The minimum of the Jeans mass is at $M \sim 10^3 $ \msolar 
in both cases and it is at this radius that we expect the first gravitational instability to 
appear and subsequently fragment and form Pop III stars \citep{Turk_2009, Clark_2011, 
Greif_2011, Greif_2012}. The atomic mass halos (1052, 1054, 1056, 1058) show qualitatively 
similar behavior but important differences exist. Simulations 1052 and 1054 both display 
gravitational instability at $M \sim 10^3 $ \msolarc, this is because the mass accretion rate 
onto these halos is higher than in both 1050 and 1051 and hence more mass has accumulated at each 
radius. The enclosed mass exceeds the Jeans value at approximately $\rm{M \sim 10^3} $ \msolarc. 
The reason for this is because they are both able to rapidly form \molH within 10 parsecs 
(reaching fractions of approximately $1 \times 10^{-5}$ at 10 parsecs), this increase in \molHc, 
and corresponding decrease in temperature, drive the Jeans value downward and 
the predominantly molecular hydrogen clump becomes self-gravitating. However, in both 1056 and 
1058, the \molH fraction does not reach the level of $1 \times 10^{-5}$ until closer to 1 parsec 
due to the higher dissociation rate. As a result the formation of a smaller clump is suppressed 
in simulations 1056 and 1058 although it is not completely negated as \molH still readily forms 
within the self-shielded core (see right hand panel of Figure \ref{J21}). \\
\indent It is also worth noting that in simulations 1054, 1056 and 1058 the enclosed mass profile 
displays a clear plateau at $M \gtrsim 10^5$ \msolarc. The plateau becomes more pronounced as the 
source flux increases, with a clean example emerging in simulation 1058.  
The plateau is the signature of a disk-like structure forming similar to what has been seen in 
atomic only simulations \citep[e.g.][]{Regan_2009, Regan_2014a}. The collapse at the 
center of the halo and the subsequent decrease in the timestep means that the evolution of the 
envelope containing $M \sim 10^5$ \msolar is effectively frozen out. Tracking the evolution
of the envelope from this point onwards is therefore very difficult due to its relatively 
long dynamical time compared to the dynamical time of the mass at the center of the collapse.
Nonetheless, it seems likely that this larger mass will collapse with a mass of close to 
$M \sim 10^5$ \msolar in all three cases (1054, 1056, 1058) with the possible exception of 
simulation 1054 where a smaller mass star ($M \sim 10^3$ \msolar) may initially form due to 
the larger \molH fraction in this case. 


\subsubsection{Comparison at Maximum Refinement}
\noindent In Figure \ref{FourPanelPlot} we have over-plotted several physical characteristics 
from the atomic halo (1052, 1054, 1056 and 1058) outputs when the simulation reaches 
the maximum level of refinement, which is 18 in this case. In the bottom left panel we plot 
the gas number density (hereafter referred to simply as the number density, unless 
explicitly stated otherwise) as a function of radius. The maximum number density reached in 
each simulation is $n \sim 3 \times 10^7 \ \rm cm^{-3}$. As we see each simulation 
asymptotes towards a cored profile in the inner regions of the plot. This core is due to the 
formation of a small disk at the very center of the collapsing halo - due to the formation of 
\molH and its subsequent collapse.\\
\indent In the top left panel we have plotted the radial velocity against the radius. The radial 
velocity shows a very strong inflow at a distance of a few hundred parsecs. This is where the 
gas is flowing rapidly into the halo and the point at which shock heating occurs (see Figure 
\ref{TemperatureTimeSeries}). Simulations 1054, 1056 and 1058 all show strong radial 
inflows at distances between a few parsecs and approximately 100 parsecs. In particular, there 
is a noticeable peak in the radial velocities of all four simulations at approximately 20 parsecs 
and another peak between 0.1 and 1.0 parsecs. In the case of simulations 1054 and 1056 this is due 
to the formation of an inner collapsing object with mass of $\rm{M} \sim 10^3$ \msolar (the core) 
and an outer collapsing object (the envelope) with a of mass $\rm{M} \sim 10^5$ \msolarc. In 
simulation 1058 the radial inflow is dominant at approximately 20 parsecs but with only a 
relatively weak radial inflow at the smaller collapsing radius. This is because of the relative 
lack of \molH in the core in this simulation and indicates that only the envelope will collapse 
with a mass of close to $M\sim 10^5 $\msolarc.\\
\indent In the top right panel we plot the ratio of the rotational velocity against the 
Keplerian (circular) velocity. The rotational velocity is calculated by computing the inertia 
tensor and then using the angular momentum vector to find the rotational velocity around the 
principal axis. This approach is detailed in \cite{Regan_2009} to which we direct the reader 
for further information. A value of $\rm{V_{Rot}/V_C} > 1$ indicates that the gas is rotationally 
supported (red dashed line). The rotational support ratio shows qualitatively the same behavior 
as the radial velocity plot. There is a peak in rotational support at a distance of approximately 
20 parsecs in each case and rotational support is achieved in all cases at this point. 
Simulations 1052 and 1056 achieve a peak in rotational support again inside 1 parsec 
consistent with the formation of a \molH clump at that radius. Simulation 1054 maintains a more 
consistent, rotationally supported, profile into the core. Simulation 1058 displays rotational 
support between approximately 10 parsecs and 100 parsecs, the ratio then dips below 1.0 inside 
10 parsecs and apart from a spike at 0.2 parsecs due to a spike in the \molH fraction and associated 
temperature decrease remains below 1.0 at all radii inside 10 parsecs. The inner core of 
simulation 1058 is therefore \emph{not} rotationally supported. The bottom panel
in Figure \ref{TempDenProjs} shows the rather unrelaxed nature of the inner core
in this simulation and provides an explanation for the anomalous behavior at small radii 
in this case. \\
\indent Finally, in the bottom right panel we plot the mass accretion rate against the radius.
The accretion rate is the instantaneous accretion rate calculated by taking the difference between
two outputs at the end of the simulation. The accretion rate in all four cases shows a noticeable 
plateau between a radius of a few parsecs and about 30 parsecs, this is due to the collapsing 
$\rm{M \sim 10^5}$ \msolar clump forming at this radius and 
accreting mass at a rate of $\sim 0.25$ \msolar $\rm{yr^{-1}}$.  This accretion rate is 
consistent with that predicted by \cite{Inayoshi_2014} and \cite{Schleicher_2013} for the 
formation of a supermassive star or possibly a quasi-star depending on the long term evolution 
of the acretion rate. Also worth 
noting is that simulation 1052 plateaus again at a radius between approximately 0.05 and 1 parsecs, 
this plateau is due to accretion onto the $\rm{M \sim 10^3}$ \msolar clump collapsing due to \molH
cooling. This plateau is not as well defined in simulations 1054, 1056 and 1058 due to the 
higher dissociation rate. \\
\indent In Figure \ref{TempDenProjs} we show density weighted projections of both the temperature 
and density for each of the atomic cooling halos (1052, 1054, 1056 and 1058) at the time the 
simulation first reaches the maximum refinement level. The projection is made along the 
angular momentum axis in each case and so is specific to each halo. This along with the fact 
that the time at which the simulation reaches the maximum refinement level is different in each 
case means that the morphology of each halo is significantly different in all cases. In each 
simulation the central object forms in the center of the cold, \molH cooled, gas. What is 
immediately clear is that the fraction of cold gas available in simulation 1052 is significantly
greater than in the 1056 and 1058. In both 1056 and 1058 it is clear that only a very small 
fraction of the gas is collapsing due to the formation of \molHc. Furthermore, both 1056 and 
1058 show the envelope (with a diameter of approximately 10 - 20 parsecs) surrounding the 
core. Simulation 1054 represents an intermediate case with a well defined envelope
and a substantial fraction of \molH within the core. The radiation in the case of 1054 
is not strong enough to penetrate all the way to the center meaning a larger fraction of
\molH is able to form.\\

\begin{figure}
\begin{center}
    \includegraphics[width=9cm]{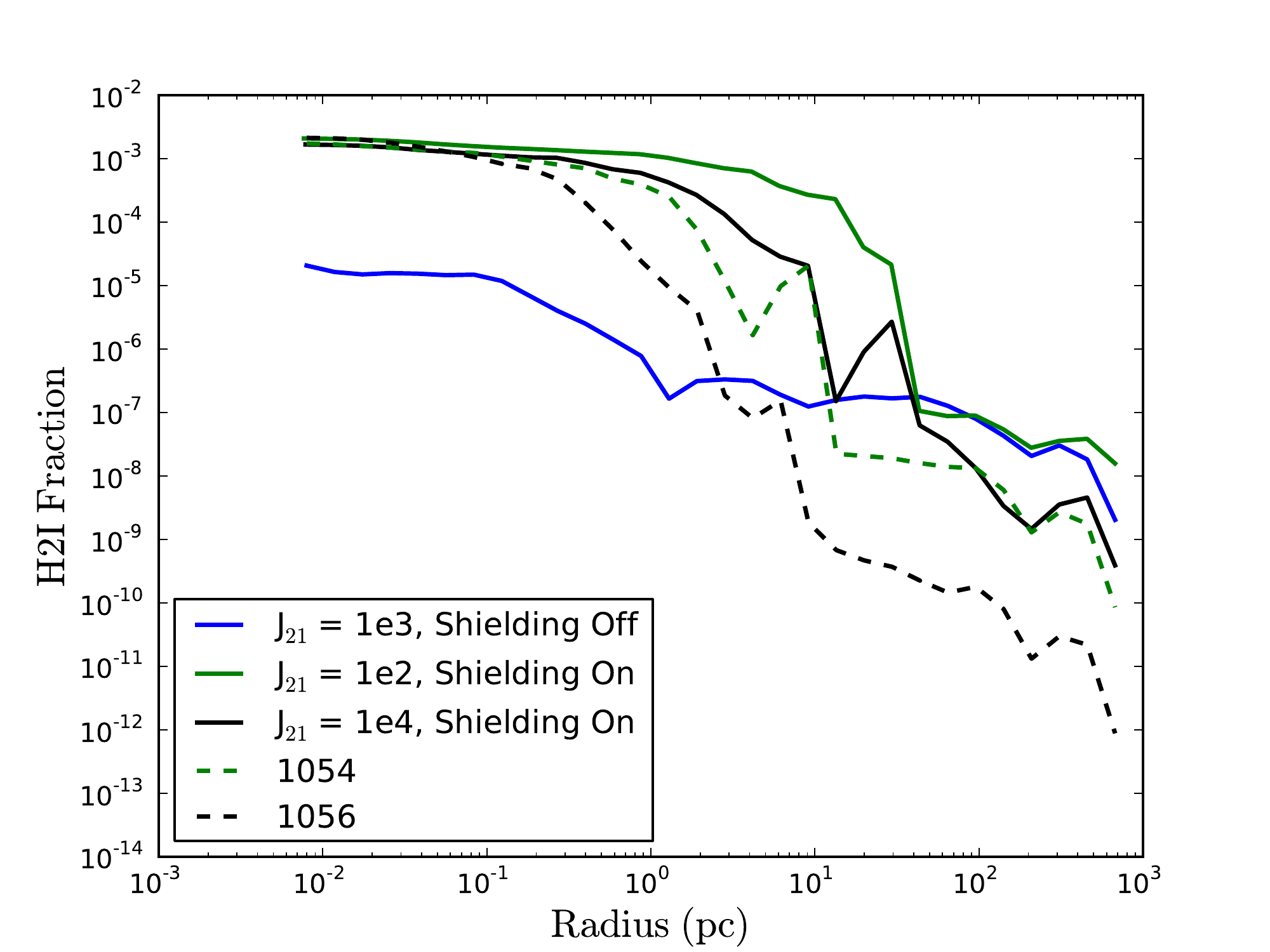}  
    \caption{\label{Isotropic}
      This figure compares the \molH fraction when the simulation is run
      using different prescriptions for the radiation. In the first three cases
      (blue, green \& black solid lines) the radiation is isotropic with different 
      intensities. Furthermore, the blue line contains no self-shielding, the 
      green and black solid lines do but the radiation intensities are different. 
      The dashed green and dashed black lines are the lines from simulations 1054 and 
      1056 for comparison. 
    }
\end{center}
\end{figure}


\begin{figure*}
    \begin{center}
      \begin{tabular}{cc}
        \includegraphics[width=16cm]{FIGURES/PDFS/Fig8b.pdf}  \\
        \includegraphics[width=16cm]{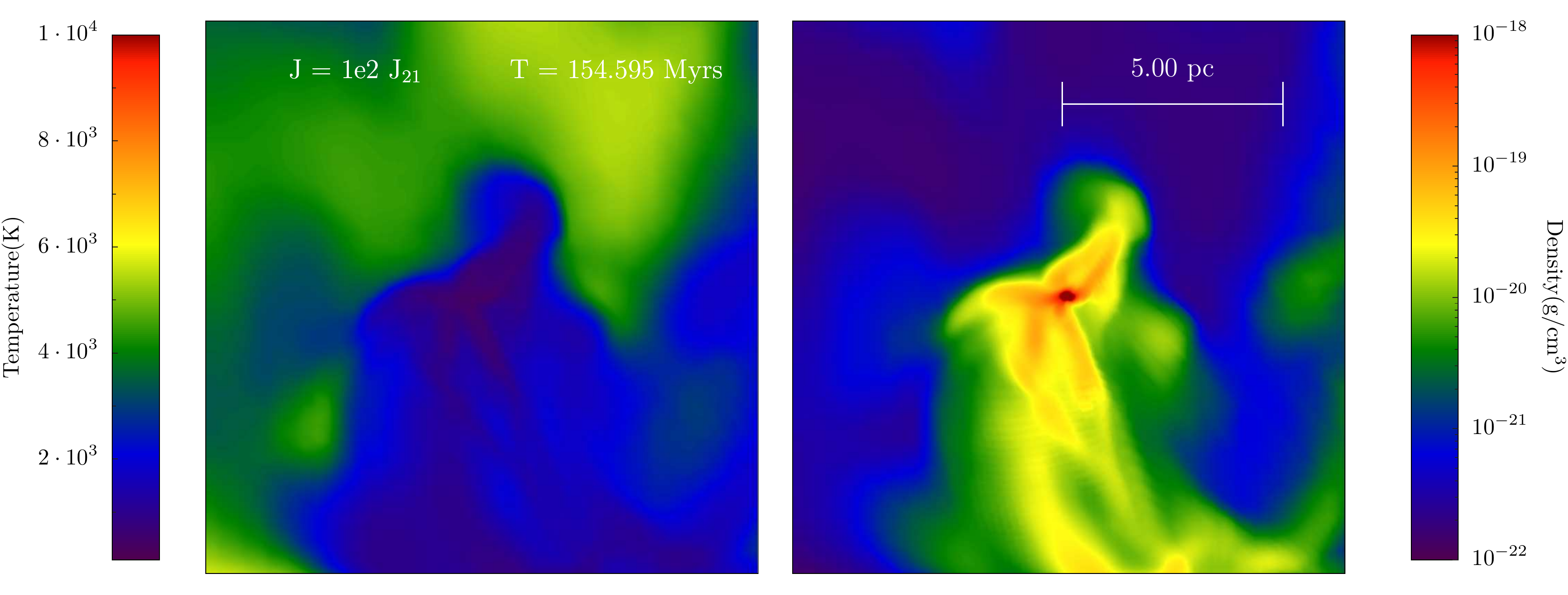}
      \end{tabular}
      
      \caption{\label{IsotropicProj}
        The top row is the same as the second row in Figure 
        \ref{TempDenProjs}, it is the projection of the 1054 simulation at the 
        highest refinement level made along the angular momentum vector. The 
        bottom row shows the projection for the $\rm{J = 1 \times 10^2\ J_{21}}$ 
        isotropic radiation case with self shielding.
        The left panel of each row shows the temperature across approximately 
        10 parsecs. The right panel in each row shows the density in the same region.
        The spatial scales of both panels are identical.
    }
    \end{center} 
\end{figure*}

\begin{figure*}
  \begin{minipage}{175mm}      
    \begin{center}
      \begin{tabular}{cc}
        \includegraphics[width=9cm]{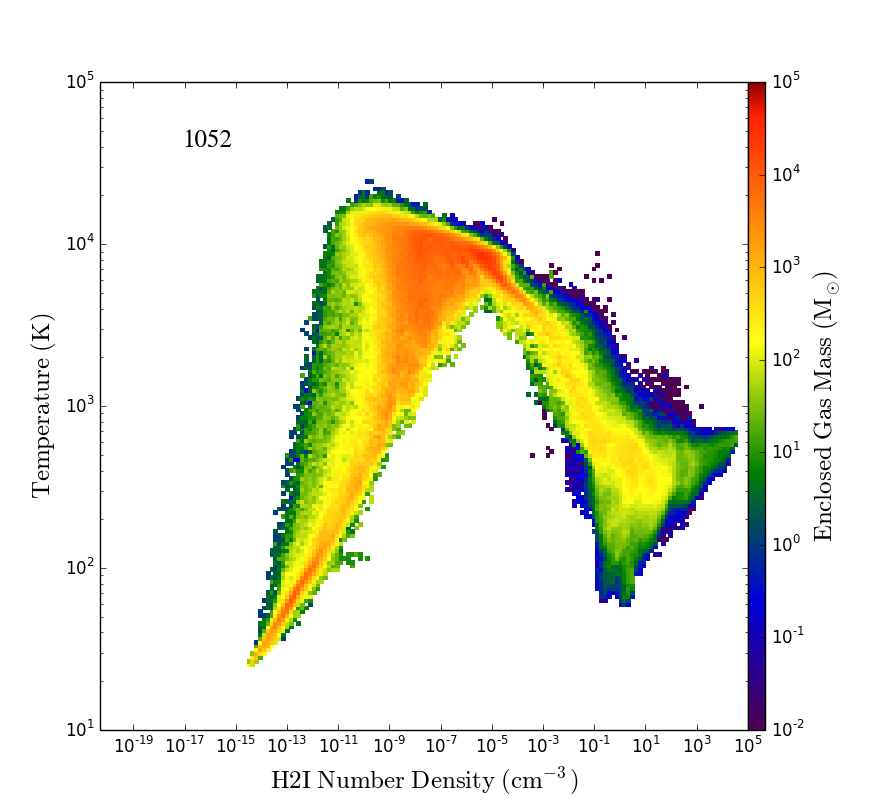}  &
        \includegraphics[width=9cm]{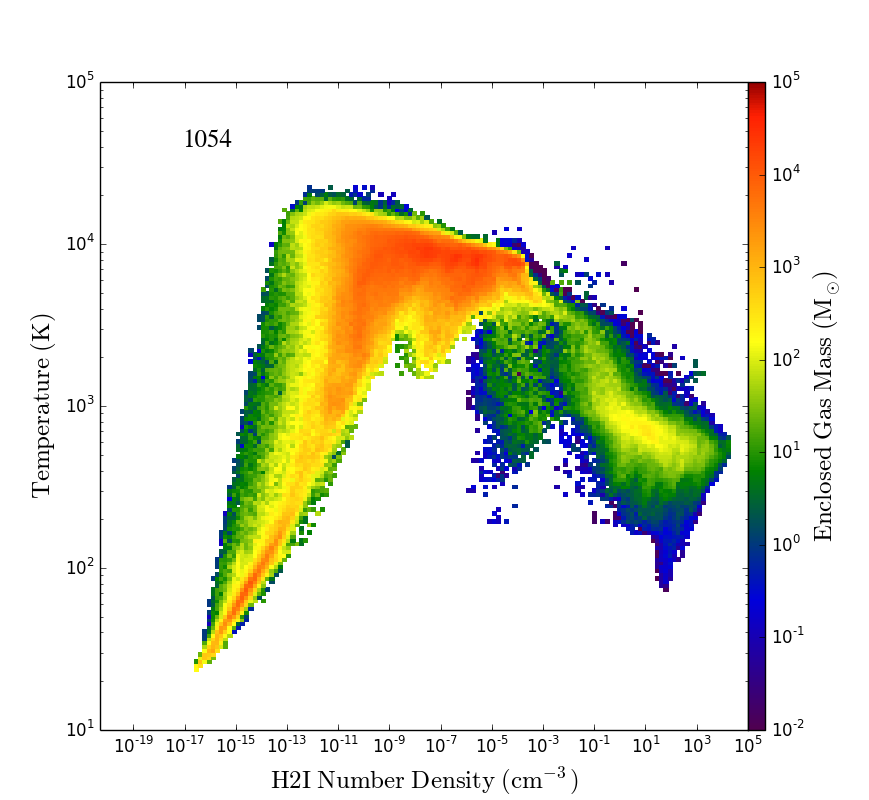}  \\
        \includegraphics[width=9cm]{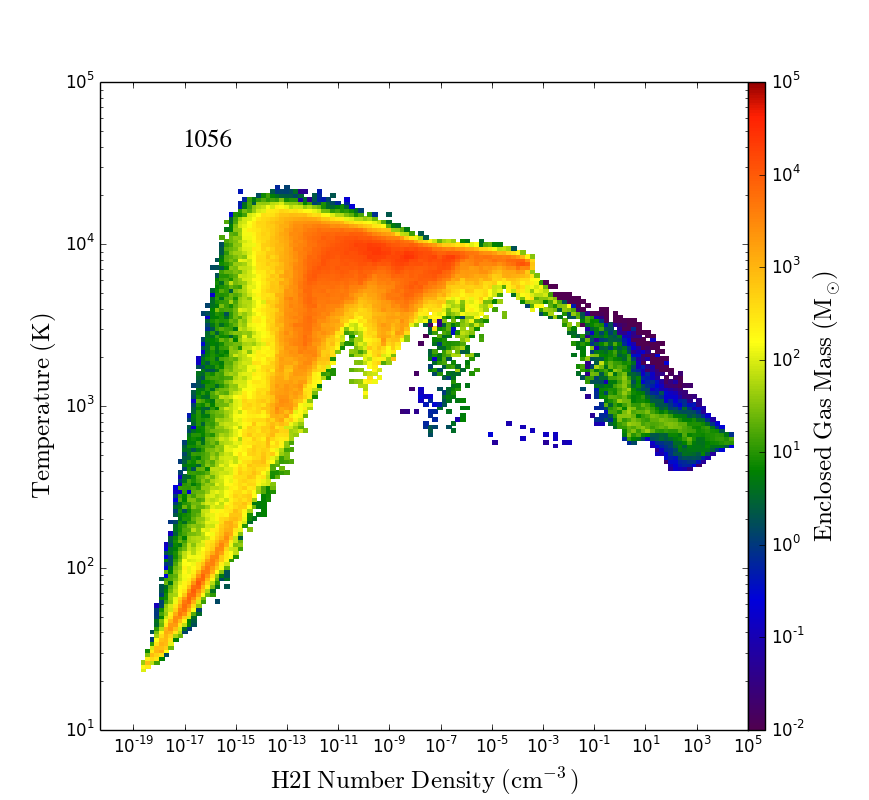}  &
        \includegraphics[width=9cm]{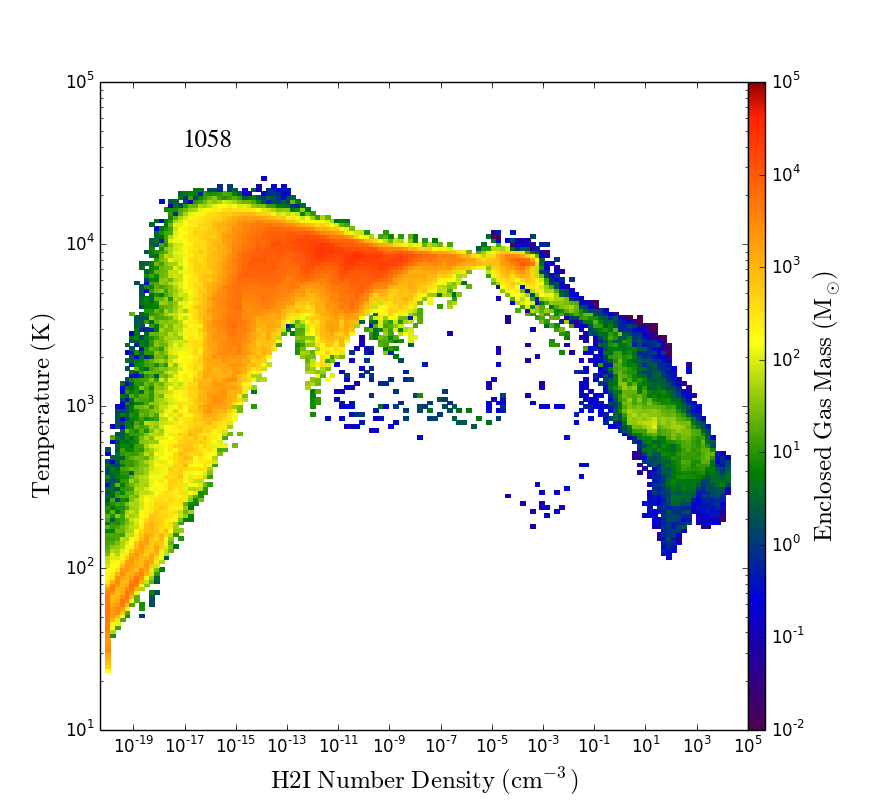}  
      \end{tabular}
      
      \caption{\label{PhasePlots}
       A Phase Space plot for each of the atomic cooling halos with \molH Number
       Density (x-axis) plotted against Temperature (y-axis). The pixels are colour coded
       according to the enclosed gas mass. In the top left panel the halo reaches atomic cooling
       status at its outer regions but within a few tens of parsecs the \molH fraction increases 
       rapidly and a significant fraction of the gas is able to cool to below $10^3$ K. As the flux 
       increases in each panel the amount of gas than is able to stay at $\sim 10^4$ K 
       increases rapidly. In the highest flux simulation (bottom right panel) only a 
       small fraction of the gas cools to below $10^3$ K. 
    }
    \end{center} 
  \end{minipage}
\end{figure*}


\section{Comparison against isotropic radiation field} \label{Sec:Isotropic}
\noindent Finally, we wish to compare, as best as possible, the effect of an isotropic 
radiation field versus an anisotropic radiation field. We make the comparisons in 
Figure \ref{Isotropic} and Figure \ref{IsotropicProj}. As done throughout this
paper outputs are compared when the simulation reaches the highest refinement level.
Fundamentally, the isotropic field means that each cell ``feels'' the same radiation 
whereas with the anisotropic source the intensity varies as $1 / r^2$ in the optically thin 
regions and according to equation \ref{DB37} in the optically thick regions. We ran a number 
of simulations, with an isotropic background and using the same halo as was used throughout 
this study in order to compare the results. Furthermore, we included a local approximation to 
calculate the \molH column density in each cell and used that value to estimate the 
self-shielding factor due to \molHc. In order to calculate the local estimate of the \molH 
column density we use the 'Sobolev-like' method \citep{Sobolev_1957, Gnedin_2009} to estimate 
the column density. In this case the characteristic length is obtained from 
\begin{equation}
L_{\rm{Sob}} \equiv {\rho \over | \nabla \rho | }
\end{equation}
where $\rho$ is the gas density. This particular method has been shown by WG11 to be particularly 
accurate in estimating the column density locally. The self-shielding approximation is then 
implemented in the same fashion as in the ray-tracing simulations, i.e. using equation 
\ref{DB37}. \\
\indent The solid blue line in Figure \ref{Isotropic} uses an isotropic background radiation 
field with $\rm{J = 1 \times 10^3~ J_{21}}$ and no \molH self-shielding. The solid green line 
uses a lower intensity field with  $\rm{J = 1 \times 10^2~ J_{21}}$ but this time \molH 
self-shielding is included. The solid black line uses a higher intensity field with 
$\rm{J = 1 \times 10^4~ J_{21}}$ and with \molH self-shielding included. For comparison the 
results from simulations 1054 (dashed green line) and 1056 (dashed black line) are also included. 
The dashed lines should therefore be compared with the solid lines of the same color.
In all cases the radiation field is switched on at a redshift of $z=32$. We specifically picked 
isotropic backgrounds with intensities of $\rm{J = 1 \times 10^2~ J_{21}}$ and 
$\rm{J = 1 \times 10^4~ J_{21}}$ so as to match closely the values simulations 1054 and 1056 
reach at the point of maximum density (i.e approximately 3 kpc from the radiation source) - 
see also left panel of Figure \ref{J21}. \\ 
\indent From Figure \ref{Isotropic} it is clear that the case with an isotropic source and 
no self-shielding is clearly different, the \molH reaches a maximum value of $\sim 2 \times 10^{-5}$ 
at the core of the halo, this differs from the self-shielded cases by approximately two orders 
of magnitude and reflects the dramatic effect of self-shielding in the core of the halo. \\
\indent In comparison the self-shielded cases all reach the \molH equilibrium 
value of $1 \times 10^{-3}$ within about 0.5 parsecs but their values at larger radii show 
significant scatter. For example, comparing simulation 1056 with the isotropic radiation field
$\rm{J = 1 \times 10^4\ J_{21}}$ (which should be approximately equal at the center of the halo), 
we see that at radii greater than about 1 parsec the \molH 
fraction differs by about two orders of magnitude. Similar differences are seen when comparing the 
isotropic field of $\rm{J = 1 \times 10^2\ J_{21}}$ with the anisotropic simulation 1054. \\
\indent For the two anisotropic cases shown in Figure \ref{Isotropic} the anisotropic source 
causes a delay in the build up of the \molH towards the center compared to the isotropic case. 
The strength of the fluxes of the anisotropic and isotropic cases were chosen to match at the 
center of the halo, however the strength of the anisotropic flux increases as one moves towards 
the radiation source compared to the isotropic case. This is why in the anisotropic case the 
\molH fraction is significantly lower in the outer parts (even though this is a radial profile
 - the destruction of \molH along the line of sight is even more extreme). The isotropic case 
has no such gradient in the flux and as a result the \molH fraction is systematically higher 
at all radii outside the core. These differences in \molH will dramatically effect 
the cooling rates at radii all the way into the core, and most pertinently within the envelope,
and hence the entire morphology and dynamics of the collapse. \\
\indent In Figure \ref{IsotropicProj} we compare in projection the central 10 pc region
of the 1054 case with the comparable isotropic field strength $\rm{J = 1 \times 10^2\ J_{21}}$.
The projections are quite clearly very different which should not be surprising given
the very different radiation field attributes in each case.

\section{Discussion} \label{Sec:Discussion}
\noindent
In this study we have looked at the effect that a single \molH dissociating source has on the 
collapse of a halo. In particular we are interested in whether a single source can keep a halo
sufficiently \molH free so that a direct collapse black hole seed may form. Using a suite of 
simulations with varying source intensities we show that while \molH is rather easily dissociated 
in the outer regions, within approximately a 10 parsec distance of the forming halo the 
\molH formation rate greatly exceeds the \molH dissociation rate and \molH forms readily. 
The formation of \molH is caused by the strong increase in the HI density as the gas collapses, 
which combines with H$^-$ to form \molH (H$^-$ + H $\rightarrow$ \molH + e).
However, as the source flux is increased the amount of \molH that can collapse is significantly 
reduced. In Figure \ref{PhasePlots} we show the \molH number density as a function of both the 
enclosed gas mass and the temperature. In simulation 1052, where the source flux is relatively 
small, the halo achieves atomic cooling status but the \molH at the center is nonetheless able 
to strongly self-shield and a large mass of \molH is able to form and collapse. This can be set 
in contrast to the situation in simulation 1058. In this case the LW flux is extremely strong, 
the halo is clearly able to collapse isothermally except for a small amount of gas at very high 
\molH density which is able to collapse and cool to below $T\sim 1000\ \rm K$. The mass of this 
cool gas $(T \sim 10^3\ \rm K)$ is of the order of 100 \msolar - at least an order of magnitude 
below the mass of cool gas that is seen in simulation 1052. \\
\indent It is clear from the preceding analysis that a very strong Lyman Werner flux is required 
to reduce significantly the ability of \molH to form. Moreover, even with a source flux 
of $1 \times 10^{58}$ photons emitted per second in the LW band (equivalent to a J of 
$\sim 1 \times 10^{7}~ \rm{J_{21}}$) \molH is still able to form within the central parsec and 
a cold clump of gas with mass $\sim 10^2$ \msolar is able to form. However, the envelope, 
with a radius of $\gtrsim 10$ parsecs is also close to gravitational instability and prone to 
collapse with a mass of about $1 \times 10^5$ \msolar resulting in the likely formation of a 
massive black hole seed.\\
\indent The final fate of the central objects subject to an anisotropic LW flux of \emph{less than}
$1 \times 10^3~ \rm{J_{21}}$  would appear to be the formation of a massive star with a 
mass of between $1 \times 10^2$ \msolar and $1 \times 10^3$ \msolar - typical of Pop III star 
formation simulations \citep[e.g.][]{Hirano_2014, Susa_2014}. As the flux, in the LW band, 
is increased to values \emph{greater than} 
$1 \times 10^3~ \rm{J_{21}}$ the mass of the collapsing object will grow as the Jeans mass 
is increased due to the declining levels of $\rm{H_2}$ and an increase in 
the mass accretion rate (see bottom right panel of Figure \ref{FourPanelPlot}). Our study 
suggests that while very strong LW fluxes from an anisotropic source cannot completely prevent the 
formation of \molH in the central regions it can reduce its impact, with the collapsing \molH 
core being very much smaller and the envelope becoming gravitationally unstable 
within a similar timescale. The collapsing total mass in this case having a mass of 
$\rm{M \sim 10^5}$ \msolarc.\\
\indent In should also be noted that the values of the fluxes shown here are
likely to be upper limits as our simulations have not included the effect of photodetachment of 
the $\rm{H^-}$ ion due to the lower energy photons. The photodetachment of $\rm{H^-}$ will remove 
the pathway for \molH formation in the core and therefore reduce the \molH formation rate. This 
of course must also be set against the effect of X-Rays and Cosmic rays \citep{Inayoshi_2011} 
which will enhance the \molH fraction. However, the detailed balance between these two feedback 
processes is unclear and will depend sensitively on an as yet undetermined Pop III initial mass 
function\citep[e.g.][]{Schneider_2006, Safranek-Shrader_2010, Hirano_2014}. Further work with a 
more comprehensive stellar spectrum and a greater sample of halos will help to further elucidate 
the issue (Regan et al. 2014c in prep).  \\
\indent Two further numerical limitations of our method concern the \molH collisional 
dissociation rates and our minimum dark matter particle mass. The \molH collisional 
dissociation rates used in this paper are those of \cite{Flower_2007}. There is some 
uncertainty in literature regarding the most appropriate collisional dissociation rate to use with 
different dissociation rates differing by up to an order of magnitude \citep{Turk_2011}. 
Our minimum dark matter particle mass is $\rm{M_{DM}} = 8.301 \times 10^2$ \msolarc. 
\enzo does not refine the dark matter particles during the collapse and so this 
exists as a numerical limitation of our method. We will investigate the limiting effects of 
both of these points in a future study.\\
\indent In \S \ref{Sec:J21} we noted that an effective stellar temperature of 
$T_{\rm eff} \sim 50000$ K is required to produce a spectrum which peaks in the LW bands. We have 
since determined that the flux in the LW band required to effectively dissociate \molH is  
\emph{greater than} $10^{54}$ photons per second. A star with an effective stellar temperature of  
$T_{\rm eff} \sim 50000$ K will produce approximately $1 \times 10^{49}$ photons per second in the 
LW band. Therefore, a galaxy with a greater than $10^5$ massive stars will be required to 
produce such a spectrum. In the early Universe, where a tilt towards a top-heavy IMF 
is expected \citep[e.g.][]{Hirano_2014, Susa_2014} such a scenario is entirely plausible 
within biased regions with close halo pairs \citep{Dijkstra_2008, Dijkstra_2014}. \\
\indent The accretion rates found in this study are $\approx 0.25$ 
\msolar $\rm{yr^{-1}}$, these rates are consistent with the accretion rates
found by \cite{Latif_2013a} and are furthermore consistent with the rates derived by 
\cite{Ferrara_2014} in which they determine the properties of the hosting haloes and the 
mass distribution function of the forming seed black holes. They find that the initial mass 
function (IMF) of the seed black holes is bimodal extending over a broad range of 
masses, M $\approx 0.5 -– 20 \times 10^5$ \msolarc. This value for the IMF is consistent 
with the value we find for the final mass of the collapsing object. \\
\indent Finally, at the high redshifts probed in this study ($z \gtrsim 20$) supersonic baryonic 
streaming velocities \citep{Tseliakhovich_2010} are a further possible mechanism which can 
affect the formation of a direct collapse black hole. Recent work by \cite{Tanaka_2014} has 
shown that relative baryonic streaming velocities may induce direct collapse black holes by 
minimizing metal enrichment and enhancing turbulent effects within a collapsing halo promoting
the creation of cold accretion filaments. The cold filaments can drive accretion rates and lead
to the formation of a direct collapse black hole.  3D hydrodynamical simulations presented by 
\cite{Latif_2014c} also supports this scenario at very high redshifts. However, 
\cite{Visbal_2014} pointed out that in such a scenario the formation of \molH is very 
difficult to prevent in the absence of a strong, nearby, LW source. Their numerical simulations 
show that streaming velocities cannot result in densities high enough to allow the halo 
to reach the ``zone of no return'' and conclude that this pathway is not a viable mechanism
for direct collapse black hole formation. We have not included 
the effect of relative streaming velocities in our calculations and based on the previous
findings we do not expect this to have a significant impact on our conclusions. 

\section{Conclusions}
\label{Sec:Conclusions}
\noindent We use the radiative transfer module in \enzo to track LW photons as they are emitted 
from a source approximately 3 kpc from a collapsing halo at very high redshift ($z\sim 30$). We 
include only the dissociating effects of this anisotropic source neglecting any effect from a 
global LW background.  We run a number of simulations with the only difference between each run 
being the LW source flux intensity. \\
\indent Our results show that as the dissociating flux is increased beyond the expected 
global average at high redshift ($\rm{J_{\rm{LW\ Global}} \approx 1 \times J_{21}}$) the collapse of 
the halo is delayed significantly and the primary means of cooling the gas is due to HI. The 
LW flux initially easily dissociates the \molH around and within the nearby, collapsing, halo. 
However, in each of our simulations, even those utilizing a very strong LW flux, \molH is 
subsequently able to form in the center of the collapsing halo due to the rapid
increase of HI (which combines with $\rm{H^-}$ to form \molH). The collapse is initiated 
by HI cooling but as the density of HI increases driving the formation of $\rm{H_2}$ 
the fraction of \molH increases rapidly in the very central regions of the collapsing object. 
The amount of \molH which is able to form is inversely proportional to the source flux intensity. \\
\indent With low source fluxes of $\sim 1 \times 10^{52}\  (\sim10^1\ \rm{J_{21}})$ photons per 
second a significant mass is able to become self-gravitating due to \molH cooling. However, 
as the source flux intensity is increased the \molH that is able to form at the center, due 
to self-shielding, decreases significantly but cannot be entirely prevented. The formation 
of $\rm{H_2}$ at the centers of the halos, even in the presence of a very strong dissociating 
LW source is therefore inevitable. \\
\indent Our study also reveals that the envelope, with a mass of
$\gtrsim 1 \times 10^5$ \msolarc, is showing strong signs of collapse over a similar timescale
as the core of \molH cooled gas. In the halos subject to a flux of \emph{greater than} 
$\sim 1 \times 10^{54}\ (\sim10^3\ \rm{J_{21}})$ photons per second, only the inner parsec 
contains significant amounts of $\rm{H_2}$ amounting to at most $\sim 100$ \msolar of $\rm{H_2}$. 
This rather small mass is surrounded by a much larger collapsing mass at higher temperatures 
(due to HI cooling). This envelope is undergoing rapid collapse at the end of the simulation, 
with accretion rates of $\sim 0.25$ \msolar $\rm{yr^{-1}}$, and is already forming a well 
defined disk with a mass of $\rm{M \sim 10^5}$ \msolar which is rotationally supported 
with a strong radial inflow. \\
\indent The formation of this rotationally supported disk is similar in 
appearance to previous work carried out where \molH cooling was either neglected 
\citep{Regan_2009, Regan_2014a} or where \molH formation was strongly suppressed
\citep{ Latif_2013c, Latif_2013a}. \cite{Regan_2014a} showed, using very high resolution 
simulations, that the envelope may fragment into star-forming clumps surrounding a central 
black hole seed. This fragmentation could lead to a dense star cluster which may undergo 
core collapse to form a massive black hole seed \citep{Davies_2011} or the clumps may 
subsequently merge with the forming protostar to form a supermassive star 
\citep{Inayoshi_2014b}. \\
\indent  Our goal was to investigate the effect of an anisotropic source flux 
on the formation of a possible black hole seed. Our simulations show that for an anisotropic
source at high redshift a rather high source intensity is required when only LW 
photons are considered. Arising from our study we can draw the following conclusions:
\begin{itemize}
\item An anisotropic source, with a LW flux $\gtrsim 1 \times 10^{54}\ (\sim10^3\ \rm{J_{21}})$
photons per second is required to suppress \molH sufficiently and allow a larger mass, 
$\rm{M \sim 10^5} $ \msolarc, 
to form. 
\item An anisotropic source, with a LW flux $\lesssim 1 \times 10^{54}\ (\sim10^3\ \rm{J_{21}})$
photons per second will form a clump of $\rm{M \sim 10^3}$ \msolar which will collapse due 
to \molH cooling and form a more typical Pop III star. 
\item A flux of $\gtrsim 1 \times 10^{52}  (\sim 10\ \rm{J_{21}})$ photons per second delays the 
collapse by up to approximately 75 Myrs compared to the case where no LW source is present. 
Stronger fluxes have little further effect on the collapse time.
\item Accretion rates of $\ge 0.2$ \msolar $\rm{yr^{-1}}$ are found for halos experiencing
strong fluxes ( $\gtrsim 1 \times 10^{54}\ (\sim10^3 \rm{J_{21}})$) photons per second in the LW band.
Accretion rates of this magnitude are ideal for the formation of either a supermassive star 
\citep{Inayoshi_2014}, a quasi-star \citep{Begelman_2008,Schleicher_2013} or a dense stellar 
cluster which subsequently undergoes core collapse \citep{Lupi_2014, Davies_2011} \\
\end{itemize}

\noindent Given that we have not included the effect of photo-detachment of $\rm{H^-}$ due to 
photons in the infrared wavelength which would also suppress \molH formation our results should 
be taken as an upper limit; however at such high redshifts, massive stars most likely dominate 
the background and relatively little infrared will be present. It therefore seems likely that a 
single strong anisotropic source, peaking at LW frequencies, with a flux \emph{greater than} $J = 
10^3~  \rm{J_{21}}$ near the collapsing halo will be sufficient to enable to the formation of 
a massive black hole seed with a mass of approximately $1 \times 10^5$ \msolarc.

\begin{acknowledgements}
\noindent J.A.R. and P.H.J. acknowledge the support of the Magnus Ehrnrooth Foundation, the Research
Funds of the University of Helsinki and the Academy of Finland grant 274931.
J.H.W. acknowledges support by NSF grants AST-1211626 and
AST-1333360.  The numerical simulations were performed on facilities 
hosted by the CSC -IT Center for Science in Espoo, Finland, which are financed by the 
Finnish ministry of education. 
Computations described in this work were performed using the publicly-available \enzo code 
(http://enzo-project.org), which is the product of a collaborative effort of many independent 
scientists from numerous institutions around the world.  Their commitment to open science has 
helped make this work possible. The freely available astrophysical analysis 
code YT \citep{YT} was used to construct numerous plots within this paper. The authors would 
like to express their gratitude to Matt Turk et al. for an excellent software package. 
J.A.R. would also like to thank Martin Haehnelt and Debora Sijacki for useful discussions 
leading to this work. Additionally, we thank Greg Byran for providing very useful comments
which helped to improve the paper. Finally, we would like to thank the referee for clear and concise
comments which further improved the overall paper. 

\end{acknowledgements}

\end{document}